\documentclass[journal]{IEEEtran}
\usepackage{algorithmic}
\usepackage[noline,linesnumbered,ruled]{algorithm2e}
\usepackage{amsmath}
\usepackage{amssymb}
\usepackage{amsfonts}
\usepackage{cite}
\usepackage{comment}
\usepackage{cases}
\usepackage{graphicx}
\usepackage{latexsym}
\usepackage[caption=false,font=footnotesize]{subfig}
\usepackage{color}
\usepackage{bm}
\newcounter{MYtempeqncnt}
\newtheorem{theorem}{Theorem}
\newtheorem{proposition}{Proposition}

\begin{document}

\title{Outage Balancing in Downlink Non-Orthogonal Multiple Access With Statistical Channel State Information}

\author{Sulong~Shi,
        Longxiang~Yang,
        and Hongbo~Zhu% <-this % stops a space
\thanks{S. Shi, L. Yang, and H. Zhu are with the College of Telecommunications and Information Engineering, Nanjing University of Posts and Telecommunications, Nanjing, 210003,
P. R. China (e-mail: \{2011010116, Yanglx, Zhuhb\}@njupt.edu.cn)}% <-this % stops a space
}

\maketitle

\begin{abstract}
This paper considers a downlink non-orthogonal multiple access (NOMA) system where the source intends to transmit independent information to the users at targeted data rates under statistical channel state information at the transmitter. The problem of outage balancing among the users is studied with the issues of power allocation, decoding order selection, and user grouping being taken into account.  Specifically, with regard to the max-min fairness criterion, we derive the optimal power allocation in closed-form and prove the corresponding optimal decoding order for the elementary downlink NOMA system. By assigning a weighting factor for each user, the analytical results can be used to evaluate the outage performance of the downlink NOMA system under various fairness constraints. Further, we investigate the case with user grouping, in which each user group can be treated as an elementary downlink NOMA system. The associated problems of power and resource allocation among different user groups are solved. The implementation complexity issue of NOMA is also considered with focus on that caused by successive interference cancellation and user grouping. The complexity and performance tradeoff is analyzed by simulations, which provides fruitful insights for the practical application of NOMA. The simulation results substantiate our analysis and show considerable performance gain of NOMA when compared with orthogonal multiple access.
\end{abstract}


\section{Introduction}
\label{sec:introduction}
\IEEEPARstart{T}{he} concept of  superposition coding (SC) is originally proposed for broadcast channels (BCs), where the disparity in the channel qualities of the users due to the near-far effect and the random fading nature of the radio channels can be exploited as a new degree of freedom for potential performance gains \cite{1972BC}. Recently, SC has received renewed attentions for its large potential in throughput and user fairness enhancement \cite{2011Tutorial,2012Experimental}. In particular, non-orthogonal multiple access (NOMA) by using SC at the transmitter and successive interference cancellation (SIC) at the receiver has been widely studied and is recognized as a promising candidate for future 5G systems. In \cite{2013MUSchedulerHM,2015MUScheduler}, the problem of multi-user scheduling in NOMA is investigated. In \cite{2013SystemLevelSimPIMRC,2014SystemLevelSimVTC} and references therein, system-level performance evaluation of NOMA is conducted from various perspectives. All these works show considerable performance gain of NOMA when compared with orthogonal multiple access (OMA).

In NOMA systems, independent signals dedicated to different users are transmitted concurrently in the same time-frequency channel, which causes strong inter-user interference when decoding at the receiver sides. To attain the potential performance gain of NOMA, advanced reception technique that can distinguish the overlapped signals is required \cite{2006OrthvsNon-orth}. SIC is the mostly considered for its low complexity and simplicity of application in existing systems \cite{2005SIC,2011Tutorial}. However, SIC gives rise to the problem of decoding order selection at the receivers, which complicates the performance optimization of NOMA, since the decoding order selection problem is of combinatorial nature and difficult to solve in many cases of interest. Most of the existing work on NOMA assume that instantaneous channel state information (CSI) is available at the transmitter \cite{2013MUSchedulerHM,2015MUScheduler,2013SystemLevelSimPIMRC, 2014SystemLevelSimVTC}. In this case, the fading BC (which models the downlink NOMA system from an information-theoretic viewpoint) can be treated as multiple parallel degraded BCs \cite{1997ParallelBC,2001FadingBCECapacity,2001FadingOutCapacity};
hence, only the decoding order selection problem of the degraded BC needs to be considered.
The capacity-achieving decoding order of a degraded BC is already very clear in the literature \cite{2005FundamentalsWirelessCom}.
However, when CSI is unavailable at the transmitter, the problem becomes rather complicated. In this case, only the outage capacity region of the fading BC is solved \cite{2009fadingBCOutCapacity, 2012fadingBCOutCapacity}. Interested readers can refer to \cite{2003FadingBCISIT,2011FadingBCISIT,2011FadingBCITA,2012FadingBCTIT} and references therein for the information-theoretic work on the fading BC without transmit CSI.

The lack of transmit CSI is a non-trivial case of practical interest. This may happen when the feedback links are limited or for other reasons. In contrast, statistical CSI is easier to obtain, especially in NOMA systems where multiple users are involved in a transmission. Hence it is improtant to investigate the performance of NOMA under statistical CSI.
We focus on the downlink scenario where the source transmits to each user at a targeted data rate, for which the outage probability is an appropriate performance metric \cite{1991OutageProbability-BlockFading,1994Blockfading-Outage}.
In \cite{2014NOMA5G}, the outage performance of the downlink NOMA system was analyzed for preassigned power allocation and decoding order while not the optimal one. The work in \cite{2015NOMA5G} investigated the impact of power allocation on the fairness outage performance of the downlink NOMA system. However, they did not consider the decoding order selection problem, only the optimal power allocation for a preassigned decoding order was solved.

In this paper, power allocation and decoding order selection are jointly studied for the downlink NOMA system to balance the outage probabilities of the users. As in \cite{2015NOMA5G}, the max-min fairness criterion is considered. Moveover, to facilitate more flexible fairness modeling, weighting factors are assigned for the users to indicate the relative importance of their success probabilities. The purpose of this work is two folds, one is to provide a theoretical tool for evaluation of the fairness outage performance of NOMA in the downlink scenario, the other is to analyze the complexity and performance tradeoff of NOMA under the investigated system setup. The fairness enhanced nature of NOMA has been extensively studied \cite{2011NOMAFairness,2012NOMAFairness}. However, these work concentrated on the achievable rate performance with the assumption that CSI is available at the transmitter, fairness with regard to the outage performance is not considered. The second purpose is motivated by the fact that the complexity of SIC scales at least linearly with the number of the users \cite{2005FundamentalsWirelessCom,2005SIC,2001SIC}, which forms the main obstacle to the practical application of NOMA. We investigate the effectiveness of user grouping in combating this problem. Specifically, we can divide the users into multiple groups and schedule only one group of the users at each time. Obviously, the complexity of SIC depends on the size of the groups. Also, the power and resource allocation among different user groups as well as the user grouping algorithm will introduce new complexity issues. All these aspects are taken into account when analyzing the complexity and performance tradeoff in our work. The main contributions of this paper are as follows:
\begin{itemize}
  \item
      %Deriving the optimal power allocation and decoding order analytically for outage balancing in the elementary downlink NOMA system where the source transmits to all active users simultaneously (i.e., without user grouping). Note that in \cite{2015NOMA5G}, the power allocation problem is studied for a given decoding order only; in addition, the optimal power allocation is obtained by using an iterative search algorithm which introduces nontrivial computational complexity.
      Solving the joint power allocation and decoding order selection problem analytically for outage balancing in the elementary downlink NOMA system where all users are scheduled simultaneously (i.e., without user grouping). Note that in \cite{2015NOMA5G}, the power allocation problem was investigated for a given decoding order only; moveover,
      only numerical solution of the optimal power allocation was obtained by using an iterative search algorithm which introduces nontrivial computational complexity.
  \item
       Investigating user grouping in downlink NOMA as well as the corresponding inter-group power and resource allocation problems. A low-complexity algorithm is provided to obtain the optimal resource allocation among the user groups to balance the outage probabilities of the users.
  \item
       Conducting and comparing the simulations with different types of resource allocation (continuous, discrete, or without) and user grouping (random, optimal, or without) to investigate the complexity and performance tradeoff of the downlink NOMA system. It is demonstrated that user grouping is an effective method in reducing the complexity of NOMA, while causes only moderate performance degradation.
\end{itemize}

The remainder of this paper is organized as follows. Section \ref{sec:systemodel} introduces the system model and the optimization problems. In section \ref{sec:solve}, the power allocation and decoding order selection problems are investigated for the elementary downlink NOMA system. The case with user grouping is considered in Section \ref{sec:group}. Simulation results are given in Section \ref{sec:sim}. Also, the complexity and performance tradeoff of NOMA is discussed. Section \ref{sec:conclusion} concludes this paper.

\emph{Notations}: Throughout this paper, $\mathbb{E}(\gamma)$ denotes the expectation of the random variable $\gamma$ and we use $\gamma|_{\alpha=\alpha_0}$ to denote the value of the variable $\gamma$ when ${\alpha=\alpha_0}$. We denote by $\mathrm{Pr}(\mathcal{O})$ the probability of the event $\mathcal{O}$ and by $\overline{\mathcal{O}}$ the complementary event of $\mathcal{O}$.

\section{System Setup and Problem Formulation}
\label{sec:systemodel}
Consider a downlink NOMA system with a source node and $K$ destination nodes or users, $U_k$, $k\in \mathcal{K} =\{1,2,\cdots,K\}$. All the nodes are equipped with a single antenna. The source has the mission of delivering mutually independent information to the users at a targeted data rate, denoted by $r_k$ for each user $U_k$, $k\in\mathcal{K}$. Unlike in conventional OMA systems where each channel block can be used by at most one user, in NOMA more than one users can be scheduled simultaneously in the same channel block. Specifically, the independent signals dedicated to the scheduled users are combined at the source using SC and then transmitted to the users. At the user sides, SIC is adopted to extract the desired information from the combined signal.
It is assumed that the channels between the source and the users undergo independent Rayleigh fadings that are constant over one channel block while vary independently from block to block (i.e., block fading). Moveover, as in \cite{2015NOMA5G}, it is assumed that perfect CSI is known to the appropriate receivers while only statistical CSI is known to the source.

We consider both the cases with and without user grouping. For the case without user grouping, all the users are scheduled simultaneously in each channel block, which is termed the elementary downlink NOMA system in this paper. In this case, the signal received by user $U_k$ can be written as
\begin{equation}
    y_k = \sqrt{H_k}h_k x + z_k, k\in\mathcal{K}\textrm{,}
\end{equation}
where $x$ is the signal transmitted by the source, $h_k$ is the normalized Rayleigh fading coefficient between the source and user $U_k$, $H_k$ is the average channel gain from the source to user $U_k$, and $z_k$ is the additive white gaussian noise with zero mean and variance $N_0$ at user $U_k$. It should be noted that the transmit signal $x$ is a superposition of $K$ independent user-dedicated signals, which goes
\begin{equation}
    x = \sum_{k\in\mathcal{K}} \sqrt{\alpha_k P}x_k,
\end{equation}
where the signal $x_k$ contains the information required by $U_k$ and satisfies $\mathbb{E}(|x_k|^2)=1$, $P$ is the short-term transmit power constraint, and $\alpha_k$ is the power allocation factor (PAF) for $U_k$ denoting the proportion of the transmit power allocated to $x_k$. For notation, use $\gamma_k = P H_k |h_k|^2/N_0$ and $\Gamma_k = \mathbb{E}(\gamma_k)$ to denote the channel SNR of user $U_k$ and its mean value, respectively. Thus, $\gamma_k$ follows an exponential distribution with parameter $1/\Gamma_k$.

When decoding at each user $U_k$, $k\in\mathcal{K}$, the desired signal $x_k$ is interfered by the other users' signals. SIC will be carried out at the users to mitigate the negative effect of the inter-user interference. The interference cancellation process is determined by the decoding order which we denote by a permutation of the user indices as $\bm{\pi}=\left\{\pi_1, \pi_2, \cdots, \pi_K\right\}$, namely, if $\pi_i=k$, then $x_k$ (or $x_{\pi_i}$) is the $i$-th user signal to be decoded. At each step of SIC, the previously decoded user signals can be regenerated by using the same channel coding and modulation as having been used by the source and then cancelled out from the received signal.
Hence when decoding the $i$-th user's signal $x_{\pi_i}$, the interference from the $j$-th user's signal $x_{\pi_j}$ with $j<i$ can be removed. At the $k$-th user $U_{\pi_k}$, the SNR associated with the decoding of the $i$-th user's signal can be given by
\begin{equation}
	\gamma_{\pi_i}^{\pi_k} = \frac{\gamma_{\pi_k}\alpha_{\pi_i}}{\gamma_{\pi_k}\alpha_I^{\pi_i}+1}, k\in{\mathcal{K}}, i\leq k,
\label{eq:SNRi_m}
\end{equation}
where $\alpha_I^{\pi_i}=\sum_{j=i+1}^{K}\alpha_{\pi_j}$ denotes the sum of the PAFs of the users whose signals are decoded later than $x_{\pi_i}$. So $U_{\pi_k}$ will fail in decoding $x_{\pi_i}$ if $\gamma_{\pi_i}^{\pi_k}<2^{r_{\pi_i}}-1 $. Note that the SNR expression in \eqref{eq:SNRi_m} is based on the assumption that $x_{\pi_j}$, $j<i$ have been successfully decoded at the $k$-th user. We have the outage event of $x_{\pi_i}$ at $U_{\pi_k}$ under decoding order $\bm\pi$ as follows
\begin{IEEEeqnarray}{rCl}
    \mathcal{O}^{\pi_k}_{\pi_i}
    &=& \left\{  \bigcup_{j\in\mathcal{K},j\leq i}\gamma_{\pi_j}^{\pi_k}< 2^{r_{\pi_j}}-1 \right\}, k\in{\mathcal{K}}, i\leq k.
\label{eq:outevent_i_at_k}
\end{IEEEeqnarray}
The outage event of the $k$-th user is simply the outage event of $x_{\pi_k}$ at user $U_{\pi_k}$, i.e., $\mathcal{O}^{\pi_k}_{\pi_{k}}$.

Here, our purpose is to find the optimal power allocation and decoding order selection to balance the outage probabilities of the users under certain fairness constraint, which is formulated as the following minimum weighted success probability maximization (MinWSP-Max) problem
\begin{IEEEeqnarray*}{rCl}
\label{eq:min-maxop}
    \max_{\alpha, \bm\pi}&\quad&\min_{k\in\mathcal{K}} \left(1-\mathrm{Pr}\left\{\mathcal{O}^{\pi_k}_{\pi_k}\right\}\right)^{w_{\pi_k}}
    \IEEEyesnumber\IEEEyessubnumber\\
    \text{s.t.}&\quad& \bm{\pi}\in\bm{\Pi},
    \IEEEyessubnumber\\
    &\quad& 0\leq\alpha_{\pi_k}\leq1, \textrm{for } k\in\mathcal{K}, \bm\pi\in\bm\Pi,
    \IEEEyessubnumber\\
    &\quad& \sum_{k\in\mathcal{K}}\alpha_{\pi_k}\leq1, \textrm{for } \bm\pi\in\bm\Pi.
    \IEEEyessubnumber
    \label{eq:sumPAFs}
\end{IEEEeqnarray*}
where $w_{\pi_k}$ is the weighting factor of the $k$-th user denoting the relative importance of its success probability and $\bm{\Pi}$ is the set of all candidate decoding orders.

When user grouping is considered, the users are divided into multiple groups and only one group of users are scheduled in each channel block. In other words, the different user groups are orthogonally multiplexed. In each group, the transmission from the source to the users follows the same way as that in an elementary downlink NOMA system. In fact, each user group can be treated as an elementary downlink NOMA system with fewer users, for which a similar problem as stated in \eqref{eq:min-maxop} can be formulated. In order to balance the outage performance of the users in different groups, inter-group power and channel resource allocation should be considered. The details on these problems are deferred to Section \ref{sec:group}.

\section{Solution of the MinWSP-Max Problem}
\label{sec:solve}
To solve the MinWSP-Max problem, a main obstacle is the decoding order selection which is nonlinear and of combinatorial nature. Since the number of the candidate decoding orders is $K!$, which can be large for even moderate values of $K$, it is difficult to acquire a knowledge on the relationship between the decoding order and the outage probabilities of the users. To circumvent this challenging issue, we first study the optimal power allocation problem under a certain assumed decoding order.
The recognitions acquired on the optimal power allocation under a given decoding order make it more easier to grasp how does the decoding order affect the fairness among the users.
Then, by induction, we derive the optimal decoding order from the outage balancing perspective.
The optimal PAFs and balanced outage probabilities of the users are obtained in closed-form by investigating the optimal power allocation under the optimal decoding order.

\subsection{Optimal power allocation under a certain decoding order}
\label{sec:PAF}
Consider any decoding order $\bm{\pi}=\{\pi_1, \pi_2, \cdots,\pi_K\}$. From \eqref{eq:outevent_i_at_k}, to decode $x_{\pi_i}$ successfully at $U_{\pi_k}$, $i\leq k$ and $k\in\mathcal{K}$, a necessary condition is
\begin{equation}
    \frac{\gamma_{\pi_k}\alpha_{\pi_i}}
    {1+\gamma_{\pi_k}\alpha_I^{\pi_i}} \geq 2^{r_{\pi_i}}-1\textrm{,}
\end{equation}
which can be rephrased as
\begin{equation}
    \gamma_{\pi_k}\left[\alpha_{\pi_i}-\left(2^{r_{\pi_i}}-1\right)\alpha_I^{\pi_i}\right] \geq 2^{r_{\pi_i}}-1\textrm{.}
    \label{eq:condition1}
\end{equation}
Obviously, when $\alpha_{\pi_i}-\left(2^{r_{\pi_i}}-1\right)\alpha_I^{\pi_i}\leq0$, the inequality in \eqref{eq:condition1} can never be satisfied, namely, $x_{\pi_i}$ can never be decoded irrespective of the channel SNR. Hence, the PAFs should be selected such that
\begin{equation}
        \alpha_{\pi_k}-(2^{r_{\pi_k}}-1)\alpha_I^{\pi_k}>0, k \in\mathcal{K},
    \label{eq:constraint1}
\end{equation}
or else the users will always be in outage, which is unwanted.

In the following, we focus on the decoding of $x_{\pi_1}$ and $x_{\pi_2}$ at an arbitrary user $U_{\pi_n}$ and assume that the PAFs of the other signals are given (thus, $\alpha_I^{\pi_2}$ is known). From \eqref{eq:outevent_i_at_k}, we have
\begin{align}
    \mathcal{O}^{\pi_n}_{\pi_1}
    &= \left\{ \gamma_{\pi_n}<\gamma_{\textrm{th}}^{\pi_1} \right\}
    \\
    \mathcal{O}^{\pi_n}_{\pi_2}
    &= \left\{ \gamma_{\pi_n}<\max\left(\gamma_{\textrm{th}}^{\pi_1}, \gamma_{\textrm{th}}^{\pi_2}\right)\right\}\textrm{,}
    \label{eq:outevent_2_at2}
\end{align}
where, for convenience, we use the notation
\begin{equation}
    \gamma_{\textrm{th}}^{\pi_k}= \frac{2^{r_{\pi_k}}-1}{\alpha_{\pi_k}-(2^{r_{\pi_k}}-1)\alpha_I^{\pi_k}}, k\in\mathcal{K}
    \label{eq:outthresh}
\end{equation}
which denotes the channel SNR required to decode $x_{\pi_k}$ given that $x_{\pi_i}$, $i<k$ have already been decoded and removed from the received signal.

To proceed, we introduce the following proposition on the optimal PAFs.
\begin{proposition}
    At the optimal power allocation, the constraint in \eqref{eq:sumPAFs} is satisfied with equality, i.e.,
    \begin{equation}
        \sum_{k\in\mathcal{K}}\alpha_{\pi_k}=1.
    \label{eq:sumto1}
    \end{equation}
\end{proposition}
\begin{IEEEproof}
When $\sum_{k\in\mathcal{K}}\alpha_{\pi_k}<1$, we can decrease $\gamma_{\textrm{th}}^{\pi_k}$, $k\in\mathcal{K}$ by scaling up all the PAFs by a factor of  $\varepsilon=1/\sum_{k\in\mathcal{K}}\alpha_{\pi_k}$, which gives rise to a lower outage probability for all the users. Hence, the optimal PAFs should satisfy \eqref{eq:sumto1}.
\end{IEEEproof}

\begin{figure}[!t]
\centering{\includegraphics[scale=0.8]{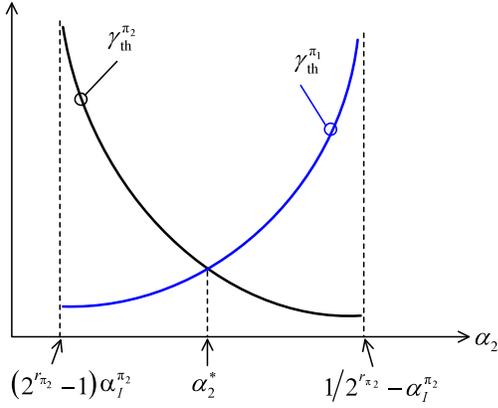}}
\centering\caption{The variation of the values of $\gamma_{\textrm{th}}^{\pi_1}$ and $\gamma_{\textrm{th}}^{\pi_{2}}$ with respect to $\alpha_{\pi_2}$.}
\label{fig:gammath1}
\end{figure}

From \eqref{eq:sumto1}, we know that $\alpha_{\pi_1}=1-\alpha_I^{\pi_2}-\alpha_{\pi_2}$. Hence, we concentrate on the selection of $\alpha_{\pi_2}$ in the following. According to $\alpha_I^{\pi_1}=\alpha_{\pi_2}+\alpha_I^{\pi_2}$ and the constraints in \eqref{eq:sumto1} and \eqref{eq:constraint1} (for $k = 1$ and $2$), we obtain the feasible range of $\alpha_{\pi_2}$ as follows
\begin{equation}
    \alpha_2\in\left((2^{r_{\pi_2}}-1)\alpha_I^{\pi_2}, 1/2^{r_{\pi_1}}-\alpha_I^{\pi_2}\right)\textrm{.}
\end{equation}
It is not difficult to see that as $\alpha_{\pi_2}$ approaches its lower bound, $\gamma_{\textrm{th}}^{\pi_2}$ turns to be infinite while $\gamma_{\textrm{th}}^{\pi_1}$ is a finite positive value, and as $\alpha_{\pi_2}$ approaches its upper bound, $\gamma_{\textrm{th}}^{\pi_1}$ turns to be infinite while $\gamma_{\textrm{th}}^{\pi_2}$ is a finite positive value. So there exists an $\alpha_{\pi_2}^*\in \left((2^{r_{\pi_2}}-1)\alpha_I^{\pi_2}, 1/2^{r_{\pi_1}}-\alpha_I^{\pi_2}\right)$ such that $\gamma_{\textrm{th}}^{\pi_1}=\gamma_{\textrm{th}}^{\pi_2}$ when $\alpha_{\pi_2}=\alpha_{\pi_2}^*$.
Then, by the monotonicity of $\gamma_{\textrm{th}}^{\pi_1}$ and $\gamma_{\textrm{th}}^{\pi_2}$ with respect to
$\alpha_{\pi_2}$, we can rephrase $\mathcal{O}_{\pi_2}^{{\pi_n}}$ in \eqref{eq:outevent_2_at2} as follows
\begin{align}
    \mathcal{O}^{\pi_n}_{\pi_2}&=
    \begin{cases}
        \left\{\gamma_{\pi_n}<\gamma_{\textrm{th}}^{\pi_1}\right\},&  \alpha_2\in\left(\alpha_2^*,1/2^{r_{\pi_1}}-\alpha_I^{\pi_2}\right),
        \\
        \left\{\gamma_{\pi_n}<\gamma_{\textrm{th}}^{\pi_2}\right\},& \alpha_2\in\left((2^{r_{\pi_2}}-1)\alpha_I^{\pi_2},\alpha_2^*\right].
    \end{cases}
\end{align}
Fig. \ref{fig:gammath1} illustrates the variation of the values of $\gamma_{\textrm{th}}^{\pi_1}$ and $\gamma_{\textrm{th}}^{\pi_2}$ with respect to $\alpha_{\pi_2}$. From Fig. \ref{fig:gammath1}, for any $\dot{\alpha}$ in $\left(\alpha_{\pi_2}^*,1/2^{r_{\pi_1}}-\alpha_I^{\pi_2}\right)$, we can find an $\ddot{\alpha}$ in $((2^{r_{\pi_2}}-1) \alpha_I^{\pi_2}, \alpha_{\pi_2}^*]$ that satisfies $\gamma_{\textrm{th}}^{\pi_2}|_{\alpha_{\pi_2}=\ddot{\alpha}}= \gamma_{\textrm{th}}^{\pi_1}|_{\alpha_{\pi_2}=\dot{\alpha}}$, namely, the required value of $\gamma_{\pi_n}$ to recover $x_{\pi_2}$ is the same when $\alpha_{\pi_2}=\ddot{\alpha}$ as when $\alpha_{\pi_2}=\dot{\alpha}$. However, the decoding of $x_{\pi_1}$ becomes easier by setting $\alpha_{\pi_2}$ to be $\ddot{\alpha}$ instead of $\dot{\alpha}$, since $\gamma_{\textrm{th}}^{\pi_1}$ decreases with $\alpha_{\pi_2}$. Thus, for the decoding of $x_{\pi_1}$ and $x_{\pi_2}$, it is optimal to select $\alpha_{\pi_2}$ in $((2^{r_{\pi_2}}-1) \alpha_I^{\pi_2}, \alpha_{\pi_2}^*]$, which is equivalent to the constraint of $\gamma_{\textrm{th}}^{\pi_1}\leq\gamma_{\textrm{th}}^{\pi_2}$. What is more, to make the decoding of $x_{\pi_1}$ and $x_{\pi_2}$ more easier is beneficial for the decoding of the following signals (i.e., $x_{\pi_i}$, $i>2$).
Hence, the optimality of the constraint $\gamma_{\textrm{th}}^{\pi_1}\leq\gamma_{\textrm{th}}^{\pi_2}$ holds true for the whole system.

The above conclusion can be extended to the selection of all the PAFs using an inductive method, for which the following theorem is given.
\begin{theorem}\label{thm:PAF}
    To achieve the optimal outage performance for a given decoding order $\bm{\pi}=(\pi_1,\pi_2,\cdots,\pi_K)$, the PAFs should satisfy the constraints in \eqref{eq:constraint1} and \eqref{eq:sumto1} and
    \begin{equation}
        \gamma_{\textrm{th}}^{\pi_1}\leq\gamma_{\textrm{th}}^{\pi_2}\leq\cdots \leq\gamma_{\textrm{th}}^{\pi_K}\textrm{.}
    \label{eq:constraint2}
    \end{equation}
\end{theorem}
\begin{IEEEproof}
    See Appendix \ref{sec:proofofTheorem1} for the proof.
\end{IEEEproof}

Theorem \ref{thm:PAF} implies that with a pre-assumed decoding order, the PAFs should be selected such that each user signal is more easier to be decoded than the user signals that are decoded later than it. From another perspective, if a user's signal can be decoded, then the signals of the users with a prior decoding order should be decodable too at the same channel SNR. Otherwise, a different decoding order from the one assumed could be used to achieve a better outage performance.

The constraints in \eqref{eq:constraint1}, \eqref{eq:sumto1}, and \eqref{eq:constraint2} form the conditions on the optimal power allocation in the elementary downlink NOMA system, which can be applied to more general outage performance analysis while not only the fairness emphasized case as considered in our work.

\subsection{Optimal decoding order from the outage balancing perspective}
Consider the decoding order $\bm\pi=\{\pi_1,\pi_2,\cdots,\pi_K\}$ and assume that the optimal power allocation conditions in \eqref{eq:constraint1}, \eqref{eq:sumto1}, and \eqref{eq:constraint2} are satisfied. Then, the weighted success probability of each user $U_{\pi_k}$ can be given as follows
\begin{align}
    \left(1-\mathrm{Pr}\left\{\mathcal{O}_{\pi_k}^{\pi_k}\right\}\right)^{w_{\pi_k}}
    =&\left(1-\mathrm{Pr}\left\{\bigcup_{j\in\mathcal{K},j\leq k}
    \gamma_{\pi_k}<\gamma_{\textrm{th}}^{\pi_j}\right\}\right)^{w_{\pi_k}}
    \nonumber\\
    =&\left(1-\mathrm{Pr}\left\{
    \gamma_{\pi_k}<\gamma_{\textrm{th}}^{\pi_k}\right\}\right)^{w_{\pi_k}}
    \nonumber\\
    =&\exp\left(-\frac{\gamma_{\mathrm{th}}^{\pi_k}}{\Gamma_{\pi_k}/w_{\pi_k}}\right), k\in\mathcal{K}
\label{eq:wsp}
\end{align}
where in the second step we have used \eqref{eq:constraint2} in Theorem \ref{thm:PAF}. Then, we have the following theorem on the optimal decoding order.
\begin{theorem}
\label{thm:order}
    For the MinWSP-Max problem in \eqref{eq:min-maxop}, there exists an optimal decoding order $\bm{\pi}^o=\{\pi_1^o,\pi_2^o,\cdots,\pi_K^o\}$, for which the following conditions are satisfied
    \begin{equation}
        \frac{\Gamma_{\pi_1^o}}{w_{\pi_1^o}}\leq\frac{\Gamma_{\pi_2^o}}{w_{\pi_2^o}} \leq\cdots\leq\frac{\Gamma_{\pi_K^o}}{w_{\pi_K^o}}.
    \end{equation}
\end{theorem}
\begin{IEEEproof}
     We consider an arbitrary decoding order $\bm{\pi}=\{\pi_1,\pi_2,\cdots,\pi_K\}$ and prove Theorem \ref{thm:order} by showing that for any two adjacent users $U_{\pi_{m}}$ and $U_{\pi_{m+1}}$, $1\leq m<K$, if $\Gamma_{\pi_m}/w_{\pi_m}>\Gamma_{\pi_{m+1}}/w_{\pi_{m+1}}$, then by exchanging the decoding orders of $x_{\pi_{m}}$ and $x_{\pi_{m+1}}$, the minimum of the weighted success probabilities of these two users can either be increased or keeped unchanged, while not affecting the weighted success probabilities of the other users. If the above statement is true, then by iteratively optimizing the decoding order of any two adjacent users, we can achieve an optimal decoding order that is the same as the one in Theorem \ref{thm:order}. See Appendix \ref{sec:proofofTheorem2} for the detailed proof.
\end{IEEEproof}

Theorem \ref{thm:order} implies that to balance the outage probabilities of the users, it is optimal to assign a higher priority in the decoding sequence for the signal of the user with a smaller weighted average channel gain, namely, \emph{the optimal decoding order depends on the ordering of the weighted average channel gains of the users}, where the weighting factor imposed on the average channel gain is the reciprocal of that imposed on the corresponding success probability. For instance, if $\Gamma_i/w_i < \Gamma_j/w_j$, then $x_i$ should be decoded prior to $x_j$.

The conclusion of Theorem \ref{thm:order} is intuitive from the weighted success probability expression in \eqref{eq:wsp} with the monotonicity of the exponential function and the optimal constraints on $\gamma_{\textrm{th}}^{\pi_k}$, $k\in\mathcal{K}$ in \eqref{eq:constraint2}.

\subsection{Optimal PAFs and the Optimized Outage Probabilities}
\label{sec:solveresult}
Suppose that the optimal decoding order as stated in Theorem \ref{thm:order} is adopted. Then, according to Theorem \ref{thm:PAF}, the MinWSP-Max problem in \eqref{eq:min-maxop} can be reformulated as
\begin{IEEEeqnarray*}{rCl}
\label{eq:min-maxop-reform1}
    \max_{\alpha} &\quad& \min_{k\in\mathcal{K}} \exp\left(-\frac{\gamma_{\textrm{th}}^{\pi_k^o}}{\Gamma_{\pi_k^o}/w_{\pi_k^o}}\right)
    \IEEEyesnumber\IEEEyessubnumber\\
    \textrm{s.t.}
    &\quad& \sum_{k\in\mathcal{K}}\alpha_{\pi_k^o}=1,
    \IEEEyessubnumber\\
    \label{eq:constraint1inminmaxop}
    &\quad& 0\leq\alpha_{\pi_k^o}\leq1, \textrm{for } k\in\mathcal{K},
    \IEEEyessubnumber\\
    &\quad& \alpha_{\pi_k^o}>\left(2^{r_{\pi_k^o}}-1\right)\alpha_I^{\pi_k^o}, \textrm{for } k\in\mathcal{K},
    \IEEEyessubnumber\\
    &\quad& \gamma_{\textrm{th}}^{\pi_1^o}\leq\gamma_{\textrm{th}}^{\pi_2^o}\leq\cdots\leq\gamma_{\textrm{th}}^{\pi_K^o}.
    \IEEEyessubnumber
    \label{eq:constraint2inminmaxop}
\end{IEEEeqnarray*}
Since $\exp(-x)$ is a monotone decreasing function of $x$, the objective function in (\ref{eq:min-maxop-reform1}a) can be replaced by $\gamma_{\textrm{th}}^{\pi_k^o}/\Gamma_{\pi_k^o}\cdot w_{\pi_k^o}$. In addition, the constraints in (\ref{eq:min-maxop-reform1}c-e) can be combined to be the following compacted form
\begin{equation} 0<\gamma_{\textrm{th}}^{\pi_1^o}\leq\gamma_{\textrm{th}}^{\pi_2^o}\leq\cdots\leq\gamma_{\textrm{th}}^{\pi_K^o}.
\label{eq:combinedconstraint}
\end{equation}
Hence, the optimization problem in \eqref{eq:min-maxop-reform1} is equivalent to
\begin{IEEEeqnarray*}{rCl}
\label{eq:min-maxop-reform2}
    \min_{\alpha} &\quad& \max_{k\in\mathcal{K}} \frac{\gamma_{\textrm{th}}^{\pi_k^o}}{\Gamma_{\pi_k^o}/w_{\pi_k^o}}
    \IEEEyesnumber\IEEEyessubnumber\\
    \textrm{s.t.}
    &\quad& \sum_{k\in\mathcal{K}}\alpha_{\pi_k^o}=1,
    \IEEEyessubnumber\\
    &\quad& \textrm{constraints in \eqref{eq:combinedconstraint}}.
    \IEEEyessubnumber
    \label{eq:s.t.}
\end{IEEEeqnarray*}
From the preconditions that $\Gamma_{\pi_1^o}/w_{\pi_1^o}\leq\Gamma_{\pi_2^o}/w_{\pi_2^o}\leq\cdots\leq\Gamma_{\pi_K^o}/w_{\pi_K^o}$ and the constraints in \eqref{eq:s.t.}, it is possible to select the PAFs such that all $\gamma_{\textrm{th}}^{\pi_k^o}/\Gamma_{\pi_k^o}\cdot w_{\pi_k^o}$, $k\in\mathcal{K}$ are equal. Inspired by this, we have the following proposition.
\begin{proposition}
\label{pro:equalWSP}
    At the optimal solution of \eqref{eq:min-maxop-reform2}, the following constraints are satisfied
    \begin{equation}
        \frac{\gamma_{\textrm{th}}^{\pi_1^o}}{\Gamma_{\pi_1^o}/w_{\pi_1^o}}=
        \frac{\gamma_{\textrm{th}}^{\pi_2^o}}{\Gamma_{\pi_2^o}/w_{\pi_2^o}}=\cdots =
        \frac{\gamma_{\textrm{th}}^{\pi_K^o}}{\Gamma_{\pi_K^o}/w_{\pi_K^o}}=A.
        \label{eq:PAFequalWSP}
    \end{equation}
    where $A$ is an auxiliary variable.
\end{proposition}
\begin{IEEEproof}
See Appendix \ref{sec:proofofProposition1} for the proof.
\end{IEEEproof}

From Proposition \ref{pro:equalWSP}, the users have equal weighted success probability at the optimal solution of the MinWSP-Max problem, which is $\exp(-A)$. The resulted outage probabilities of the users are  $P_{\pi_k^o}^{\textrm{out}}=1-\exp(A/w_{\pi_k^o})$, $k\in\mathcal{K}$. As can be expected, the user with a larger $w$ will suffer a smaller outage probability. The optimal power allocation is given in terms of the PAFs in the following theorem.
\begin{figure*}[!t]
\normalsize
\setcounter{MYtempeqncnt}{\value{equation}}
\setcounter{equation}{21}
\begin{multline}
 \qquad\alpha_{\pi_k^o} =\left\{
 \begin{IEEEeqnarraybox}[][c]{l?s}
  \IEEEstrut
   \left(\frac{1}{\Gamma_{\pi_k^o}/w_{\pi_k^o}} + \frac{2^{r_{\pi_{k+1}^o}}-1}{\Gamma_{\pi_{k+1}^o}/w_{\pi_{k+1}^o}}
    	 +\sum_{j=k+2}^{K}\frac{2^{r_{\pi_j^o}}-1}{\Gamma_{\pi_j^o}/w_{\pi_j^o}}\cdot2^{\sum_{s=k+1}^{j-1}r_{\pi_s^o}}
    	\right)\frac{2^{r_{\pi_k^o}}-1}{A} & , for $k\leq K-2, k\in\mathcal{K}$, \\
   \left(\frac{1}{\Gamma_{\pi_{K-1}^o}/w_{\pi_{K-1}^o}}+\frac{2^{r_{\pi_K^o}-1}}{\Gamma_{\pi_K^o}/w_{\pi_K^o}}\right)			 \frac{2^{r_{\pi_{K-1}^o}}-1}{A} & , for $k=K-1$, \\
   \frac{2^{r_{\pi_K^o}}-1}{\Gamma_{\pi_K^o}/w_{\pi_K^o} A} & , for $k=K$.
  \IEEEstrut
 \end{IEEEeqnarraybox}
 \right.
 \label{eq:PAFK}
\end{multline}
\setcounter{equation}{\value{MYtempeqncnt}}
\hrulefill
\vspace*{0pt}
\end{figure*}
\addtocounter{equation}{1}
\begin{theorem}
	For the elementary downlink NOMA system, to achieve the optimal balanced outage performance, the PAFs of the users should be selected according to \eqref{eq:PAFK}; on top of the next page,
where
\begin{align}
    A &=\frac{2^{r_{\pi_1^o}}-1}{\Gamma_{\pi_1^o}/w_{\pi_1^o}}+\sum_{k=2}^{K}\frac{2^{r_{\pi_k^o}}-1}{\Gamma_{\pi_k^o}/w_{\pi_k^o}}\cdot2^{\sum_{j=1}^{k-1}r_{\pi_j^o}},
    \label{eq:A}
\end{align}
\end{theorem}
\begin{IEEEproof}
First, expand and rephrase the equations in \eqref{eq:PAFequalWSP} as follows
\begin{equation}
    \alpha_{\pi_k^o}{A} = \left(\frac{1}{\Gamma_{\pi_k^o}/w_{\pi_k^o}}
    +\sum_{j=k+1}^K\alpha_{\pi_j^o}{A}\right)\left(2^{r_{\pi_k^o}}-1\right), k\in\mathcal{K}\textrm{.}
\end{equation}
Then, from the fact that $A = \sum_{k\in\mathcal{K}}\alpha_{\pi_k^o}A$, the expression of $A$ in \eqref{eq:A} can be obtained through some manipulations. The optimal PAFs can be  obtained by solving the equations in \eqref{eq:PAFequalWSP} successively from large to small with respective to the value of $k$.
\end{IEEEproof}

Recall that $\Gamma_{\pi_k^o} = P/N_0\mathbb{E}(|h_{\pi_k^o}|^2)$, from the expression of $\alpha_{\pi_k^o}$, $k\in\mathcal{K}$, the optimal PAFs are determined by the targeted data rates, average channel gains, and weighting factors of the users, while not affect by the transmit SNR, $P/N_0$. Moreover, to increase $P/N_0$ only decreases the outage probabilities of all the users by the same amount in the logarithmic scale and does not affect the relative outage performance of the users.

\section{NOMA with User Grouping}
\label{sec:group}
In the elementary downlink NOMA system, all active users are scheduled simultaneously in each channel block. However, this may be impractical, because the complexity of SIC scales at least linearly with the number of the users that are involved in a transmission\cite{2005FundamentalsWirelessCom,2005SIC,2001SIC}. In this section, we  consider the case with user grouping, in which the number of the users involved in a transmission is much smaller than $K$,  and hence a better  complexity and performance tradeoff can be achieved. The associated problems of power and resource allocation among different user groups are investigated from the outage balancing perspective.

\subsection{User Grouping}
To realize NOMA with user grouping, the first problem is how to group the users. We focus on the case when all the user groups have the same number of users, denoted by $L$. This is considered because it is simple to be realized and (as will be shown) can achieve a good enough performance.
Use $\mathcal{G}=\{1,2,\cdots,G\}$ to denote the set of the user groups, where $G=K/L$.
Though it is assumed that $K$ is a multiple of $L$, the following analysis can be applied to all possible values of $K$. Specifically, when $K$ is not a multiple of $L$, we can add $GL-K$ virtual users with extremely good channel qualities  (or extremely small weighting factor)  to the user set.
The system with virtual users can well approximate the original one, since the power required by the virtual users is nearly zero, which has trivial effect on the performance of the other users.

Two grouping algorithms will be considered, random grouping and optimal grouping. At each time, the random grouping algorithm randomly selects $L$ users from the ungrouped user set and forms them into a new group, which is repeated until there are no users remained.
The computational complexity of random grouping is $O(K)$.
The optimal grouping algorithm is realized by recursive search of all candidate grouping modes. The number of the grouping modes is ${K!}/{\big((L!)^{\frac{K}{L}}\frac{K}{L}!\big)}$, which increases exponentially with $K$.

\subsection{Inter-Group Power and Resource Allocation}
For downlink NOMA with user grouping, each user group can be treated as an elementary downlink NOMA system with $L$ users. Similarly as in Section \ref{sec:solveresult}, there should be an ``$A$" for each user group, which we denote by $A_g$ for $g\in\mathcal{G}$. Use $p_g$ and $t_g$ respectively to denote the proportion of the power and the channel resource allocated to group $g$. Then according to \eqref{eq:A}, $A_g$ can be given as follows
\begin{align}
    A_g &= \frac{2^{r_{g_1}/t_{g}}-1}{p_{g}/t_{g}\Gamma_{g_1}/w_{g_1}} + \sum_{l=2}^L \frac{2^{r_{g_l}/t_{g}}-1}{p_{g}/t_{g}\Gamma_{g_l}/w_{g_l}}2^{\sum_{j=1}^{l-1}r_{g_j}/t_{g}}
    \nonumber\\
    &=f_g(t_g)/p_g
    \label{eq:Ag}
\end{align}
where $g_l$ is the index of the $l$-th user in group $g$ after sorting the $L$ users according to
Theorem \ref{thm:order}, i.e., $\Gamma_{g_1}/w_{g_1}\leq\Gamma_{g_2}/w_{g_2} \leq\cdots\leq\Gamma_{g_L}/w_{g_L}$
and
\begin{equation}
	f_g(t_g) = t_{g}\left(\frac{2^{r_{g_1}/t_{g}}-1}{\Gamma_{g_1}/w_{g_1}} + \sum_{l=2}^L \frac{2^{r_{g_l}/t_{g}}-1}{\Gamma_{g_l}/w_{g_l}}2^{\sum_{j=1}^{l-1}r_{g_j}/t_{g}}\right).
\end{equation}
The corresponding optimal PAFs of the $L$ users in group $g$ can be obtained according to \eqref{eq:PAFK}, which are not shown here for save of space.

Note that the $L$ users in each group $g$, $g\in\mathcal{G}$ will have the same weighted success probability which monotonously decreases with $A_g$. Hence, the problem of inter-group power and resource allocation to maximize the minimum weighted success probability of the $K$ users is simply to minimize the maximum $A_g$, $g\in\mathcal{G}$, which can be formulated as follows
 \begin{IEEEeqnarray*}{rCl}
\label{eq:min-maxAg}
    \min_{p_g,t_g, g\in\mathcal{G}} &\quad& \max_{g\in\mathcal{G}} A_g
    \IEEEyesnumber\IEEEyessubnumber
    \\
    \textrm{s.t.}
    &\quad& p_g\geq0, g\in\mathcal{G},
    \IEEEyessubnumber
    \label{eq:p_g>=0}
    \\
    &\quad& \sum_{g\in\mathcal{G}}p_g\leq1,
    \IEEEyessubnumber
    \label{eq:sum_p_g}
    \\
    &\quad& t_g\geq0, g\in\mathcal{G},
    \IEEEyessubnumber
    \label{eq:t_g>=0}
    \\
    &\quad& \sum_{g\in\mathcal{G}}t_g\leq1.
    \IEEEyessubnumber
    \label{eq:sum_t_g}
\end{IEEEeqnarray*}
For the above optimization problem, we have the following propositions
\begin{proposition}
	At the optimal solution of \eqref{eq:min-maxAg}, the following conditions must be satisfied
	\begin{equation}
		A_g=A_0, g\in\mathcal{G}.
		\label{eq:Ag=}
	\end{equation}
	where $A_0$ is an auxiliary variable.
\label{pro:Ag=}
\end{proposition}
\begin{IEEEproof}
	We prove this by contradiction. Suppose that the conditions in \eqref{eq:Ag=} are not satisfied at the optimal solution and denote by $\tilde{\mathcal{G}}=\arg \max_{g\in\mathcal{G}} A_g$. It is straightforward from the monotonicity of $A_g$ with respect to $p_g$ that by decreasing all $p_g$, $g\in\mathcal{G}\setminus\tilde{\mathcal{G}}$ by an appropriate amount $\delta$ and increasing $p_{g}$, $g\in\tilde{\mathcal{G}}$ by $(G/\tilde{G}-1)\delta$, the value of $\max_{g\in\mathcal{G}} A_g$ can be decreased, where $\tilde{G}$ is the cardinality of $\tilde{\mathcal{G}}$. The proposition is proved.
\end{IEEEproof}
\begin{proposition}
	At the optimal solution of \eqref{eq:min-maxAg}, the constraints in \eqref{eq:p_g>=0} and \eqref{eq:t_g>=0} are satisfied with inequality and the constraint in \eqref{eq:sum_p_g} with equality.
\label{pro:constraints}
\end{proposition}
\begin{IEEEproof}
First, consider the constraints in \eqref{eq:p_g>=0} and \eqref{eq:t_g>=0}, if any of them are satisfied with equality, the corresponding group of users will have a zero weighted success probability, which violates the outage balancing criterion. Then, if \eqref{eq:sum_p_g} is satisfied with inequality, by scaling up  all $p_g$, $g\in\mathcal{G}$ by a factor of $1/\sum_{g\in\mathcal{G}}p_g$, all $A_g$, $g\in\mathcal{G}$ will be decreased, as $A_g$, $g\in\mathcal{G}$ motonously increases with $p_g$ when $p_g>0$, which violates the minimization criterion.
Hence, to achieve the optimal solution of \eqref{eq:min-maxAg}, the statements in Proposition \ref{pro:constraints} should be satisfied. The motonicity of $A_g$ with respect to $p_g$ is straightforward from \eqref{eq:Ag}.
\end{IEEEproof}

Proposition \ref{pro:Ag=} implies that the $K$ users will have the same weighted success probability  at the optimal inter-group power allocation, no matter how the channel resources are allocated.
From proposition \ref{pro:constraints}, the constraint in \eqref{eq:sum_p_g} is satisfied with equality at the optimal inter-group power allocation. Then, according to \eqref{eq:Ag} and \eqref{eq:Ag=}, we have
\begin{align}
    A_0 &= A_0\sum_{g\in\mathcal{G}} p_g=\sum_{g\in\mathcal{G}} p_gA_g=\sum_{g\in\mathcal{G}}f_g(t_g).
        \label{eq:A0}
\end{align}
The corresponding optimal $p_g$, $g\in\mathcal{G}$ can be obtained as
\begin{equation}
    p_g = \frac{f_g(t_g)}{A_0}.
    \label{eq:p_g}
\end{equation}

Now, we consider the optimization of the resource allocation parameters, which is further formulated as follows
\begin{IEEEeqnarray*}{rCl}
\label{eq:minA0}
    \min&\quad&A_0
    \IEEEyesnumber\IEEEyessubnumber\\
    s.t.&\quad&\sum_{g\in\mathcal{G}} t_g\leq1,
    \IEEEyessubnumber
    \label{eq:sum_t_g<=1}
    \\
    &\quad& t_g>0,g\in\mathcal{G}.
    \IEEEyessubnumber
    \label{eq:t_g>0}
\end{IEEEeqnarray*}
where the constraints in \eqref{eq:t_g>0} are based on Proposition \ref{pro:constraints}.  To solve problem \eqref{eq:minA0}, we have the following propositions
\begin{proposition}
\label{pro:convexityfg}
	For any $g\in\mathcal{G}$,  in the region of $\{t_g|t_g>0\}$, the first derivative of $f_g(t_g)$, given in \eqref{eq:fgderivate}, is a strictly monotone increasing function of $t_g$, and $f_g(t_g)$ is a strictly convex function of $t_g$.
\begin{align}
    f_g'(t_g)=
    &-\frac{w_{g_1}}{\Gamma_{g_1}}
    -\sum_{l=1}^{L-1}\left(\frac{w_{g_l}}{\Gamma_{g_l}}-\frac{w_{g_{l+1}}}{\Gamma_{g_{l+1}}}\right)
    \left(x_g^l\ln x_g^l - x_g^l\right)
    \nonumber\\
    &-\frac{w_{g_L}}{\Gamma_{g_L}}\left(x_g^L \ln x_g^L - x_g^L\right), g \in\mathcal{G}
    \label{eq:fgderivate}
\end{align}
where for convenience, we use the notation $x_g^l = 2^{\sum_{j=1}^l r_{g_j}/t_{g}}$.
\end{proposition}
\begin{IEEEproof}
	See Appendix \ref{sec:proofofconvexity} for the proof.
\end{IEEEproof}
\begin{proposition}
	Problem \eqref{eq:minA0} is strictly convex.
\end{proposition}
\begin{IEEEproof}
	Base on the convexity of $f_g(t_g)$ (see Proposition \ref{pro:convexityfg}), it is obvious that $A_0$ is a strictly convex function of $t_g, g\in\mathcal{G}$ in the region defined by \eqref{eq:t_g>0}, as $A_0$ is a linear combination of $f_g(t_g), g\in\mathcal{G}$. Moveover, the constraints in \eqref{eq:sum_t_g<=1} and \eqref{eq:t_g>0} are linear, i.e., convex. Hence,  problem \eqref{eq:minA0} is convex.
\end{IEEEproof}

We resort to the method of Lagrange multiplier to solve problem \eqref{eq:minA0}. The Lagrange function is given by
\begin{align}
    \Lambda(t_1,\cdots,t_G,\lambda)
    =  \sum_{g\in\mathcal{G}} f_g(t_g)+\lambda\left(\sum_{g\in\mathcal{G}} t_g-1\right)
\end{align}
where $\lambda$ is the Lagrange multiplier for constraint \eqref{eq:sum_t_g<=1}, and the Karush-Kuhn-Tucker (KKT) optimality conditions are as follows
\begin{IEEEeqnarray*}{rCl}
\label{eq:KKT}
    \frac{\partial \Lambda}{\partial t_g} =f_g'(t_g)+\lambda&=& 0, \quad g \in\mathcal{G},
    \IEEEyesnumber\IEEEyessubnumber
    \label{eq:f+lambda}
    \\
    \sum_{g\in\mathcal{G}}t_g&\leq&1
    \IEEEyessubnumber
    \label{eq:sum_t_g<=1KKT}
    \\
    \lambda\left(\sum_{g\in\mathcal{G}}t_g-1\right) &=& 0,
    \IEEEyessubnumber
    \label{eq:lambdatimes}
    \\
    \lambda&\geq&0,
    \IEEEyessubnumber
    \\
    t_g&>&0, \quad g\in\mathcal{G}.
    \IEEEyessubnumber
\end{IEEEeqnarray*}
Due to complicated expression of $f_g'(t_g)$, it is hard to solve the optimality equations in \eqref{eq:KKT} analytically. Here, we provide a simple iterative algorithm to obtained the optimal $t_g$'s, for which the following proposition is essential.
\begin{proposition}
	At the optimal solution of \eqref{eq:minA0}, the constraint in \eqref{eq:sum_t_g<=1} is satisfied with equality.
\end{proposition}
\begin{IEEEproof}
	Recall that problem \eqref{eq:minA0} is convex, hence the conditions in \eqref{eq:KKT} are necessary. From \eqref{eq:fgderivate}, $f_g'(t_g)<0$ always holds true for any $g\in\mathcal{G}$ in the feasible range defined by \eqref{eq:t_g>0}. Hence, by \eqref{eq:f+lambda}, $\lambda$ has to be strictly positive. Further, \eqref{eq:lambdatimes} implies that constraint \eqref{eq:sum_t_g<=1KKT} or \eqref{eq:sum_t_g<=1} must be satisfied with equality.
\end{IEEEproof}

Since for any $g\in\mathcal{G}$, $f_g'(t_g)<0$ and $f_g'(t_g)$ strictly increases with $t_g$ in the feasible range (see Proposition \ref{pro:convexityfg}), there exists a unique and strict positive solution for all equations in \eqref{eq:f+lambda}  for any given $\lambda>0$. With this observation,
the problem of finding the optimal $t_g$'s can be alternatively solved by finding the $\lambda$ which yields $t_g$'s that satisfy \eqref{eq:sum_t_g<=1} with equality. Though it is hard to find $t_g$'s adding up exactly to one, we can iteratively bound $\lambda$ to mitigate the gap of $\sum_{g\in\mathcal{G}}t_g$ to the upper bound to an acceptable value. To realize this, we have the following algorithm
\begin{description}
  \item[Step 1:]~Set $\lambda_{L}=0$, $\lambda_{H}=-\max_{g\in\mathcal{G}}f_g(1)$, $\theta=0$;
  \item[Step 2:]~Set $\lambda=(\lambda_{L}+\lambda_{U})/2$ and solve \eqref{eq:f+lambda} to obtain $t_g$, $g\in\mathcal{G}$;
  \item[Step 3:]~If $\sum_{g\in\mathcal{G}} t_g\leq1$, $\theta=0$, else $\theta=1$;
  \item[Step 4:]~If $\lambda_H-\lambda_L\leq\epsilon_o$, $t_g^*=t_g$ for $g\in\mathcal{G}$;
  \\
  			else, set $\lambda_L=(1-\theta)\lambda_L+\theta\lambda$ and $\lambda_H=\theta\lambda_H+(1-\theta)\lambda$, and go to Step 2.
\end{description}
Note that from Step 2 to 4,  many other line search algorithms can be used. In step 2, the equations in \eqref{eq:f+lambda} can be solved using the Newton's method or many others \cite{2007PracticalOptimization}. The computational complexity of the algorithm depends on the precision required. Use $\epsilon_i$ to denote the precision required when solving \eqref{eq:f+lambda}, then the total complexity is $G\log\epsilon_o\log\epsilon_i$. If parallel computing is available, the complexity will be  $\log\epsilon_o\log\epsilon_i$.

The optimal $A_0$ and $p_g$'s can be obtained by substituting $t_g^*$'s into \eqref{eq:A0} and \eqref{eq:p_g}, respectively.

\section{simulation results and discussion}
\label{sec:sim}
In this section, we present simulation results to verify our analysis and show the potential performance gain of NOMA over OMA. The tradeoff between the complexity and performance of NOMA is also discussed. We consider a network model where the users are uniformly distributed over a circular disk (with its radius being normalized to be 1 meter) centered at the source. We assume that $H_k={d_k}^{-\eta}$ with $d_k$ being the distance from the source to user $U_k$ and $\eta$ the pathloss attenuation factor which is set to be $3.75$. The other parameters, including the targeted data rates and weighting factors of the users, will be specified for each of the following results. For convenience, we use the notations $\bm r = \{r_1,r_2,\cdots,r_K\}$, $\bm w = \{w_1,w_2,\cdots,w_K\}$, and $\bm d =\{d_1,d_2,\cdots,d_K\}$. As a reference scheme, we use the conventional TDMA which refers to the orthogonal allocation of the channel resources and is essentially equivalent to any OMA scheme \cite[Sec. 6.1.3]{2005FundamentalsWirelessCom}; both the cases with only power allocation (PA) and with joint power and resource allocation (PARA) are considered. The optimal resource allocation in a TDMA system can be solved by using a methodology similar to that for inter-group resource allocation in NOMA (see Section \ref{sec:group}).

\subsection{Verification of the optimal power allocation and decoding order selection}
\begin{table}
  \caption{Randomly generated $\bm r$, $\bm w$, and $\bm d$ for the results in Fig. \ref{fig:verification}.}
  \centering
  \begin{tabular}{|c|c|c|c|}
    \hline
        & $\bm r$ (bits/s/Hz) & $\bm w$ & $\bm d$ (meter) \\
    \hline
    G1 & (0.57, 0.04, 1.39) & (0.39, 0.30, 0.31) & (0.453, 0.788, 0.417)\\
    \hline
    G2 & (0.91, 0.35, 0.74) & (0.11, 0.29, 0.60) & (0.535, 0.981, 0.480)\\
    \hline
    G3 & (0.47, 0.74, 0.79) & (0.25, 0.35, 0.40) & (0.904, 0.842, 0.208)\\
    \hline
    G4 & (0.65, 0.23, 1.12) & (0.45, 0.46, 0.09) & (0.636, 0.550, 0.870)\\
    \hline
    G5 & (0.73, 0.22, 1.05) & (0.32, 0.26, 0.42) & (0.951, 0.531, 0.784)\\
    \hline
  \end{tabular}
  \label{table:exhaustivePara}
\end{table}
\begin{figure}[!t]
\centering{\includegraphics[scale=0.645]{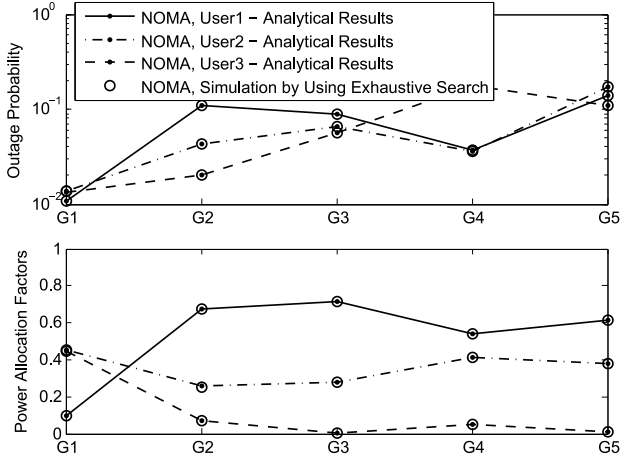}}
\centering\caption{%Verification of the analytical results by using exhaustive search for
Outage probabilities and PAFs of the users using both the analytical results and exhaustive search under five group of randomly generated simulation parameters as given in Table \ref{table:exhaustivePara}. The case of $K=3$ is considered, the sum rate is set to be $\sum_{k=1}^K r_k = 2$ bits/s/Hz, and the transmit SNR is $P/N_0=10$ dB.}
\label{fig:verification}
\end{figure}
To substantiate our analytical results, we provide the solution of the MinWSP-Max problem
by using exhaustive search of the optimal decoding order and PAFs in Fig. \ref{fig:verification}, where we take the case of $K=3$ as an example, and randomly generate five group (G1 to G5) of simulation parameters, which are given in Table \ref{table:exhaustivePara}. As can be observed from Fig. \ref{fig:verification}, the results by using exhaustive search well consolidate our analytical results, in terms of both the outage probabilities and the PAFs of the users. Also, as analyzed, the user with a larger weighting factor will have a smaller outage probability.

\begin{figure}[!t]
\centering{\includegraphics[scale=0.65]{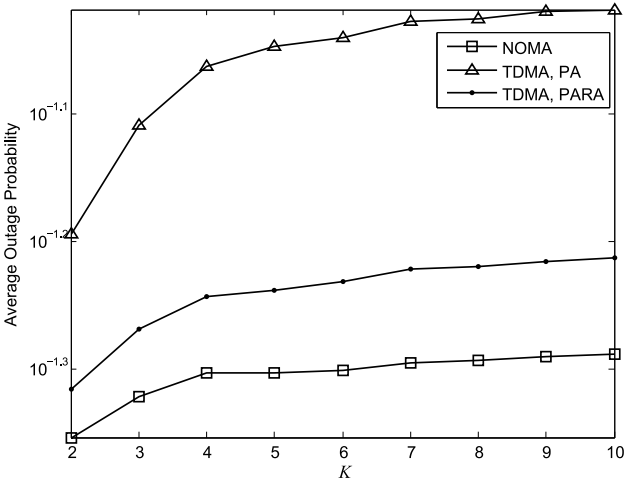}}
\centering\caption{The average outage probability of the user $1$ with respect to $K$ when $P/N_0=15$ dB and $r_{\Sigma}=3$ bits/s/Hz.}
\label{fig:w.r.t.userNum}
\end{figure}

\begin{figure}[!t]
\centering{\includegraphics[scale=0.65]{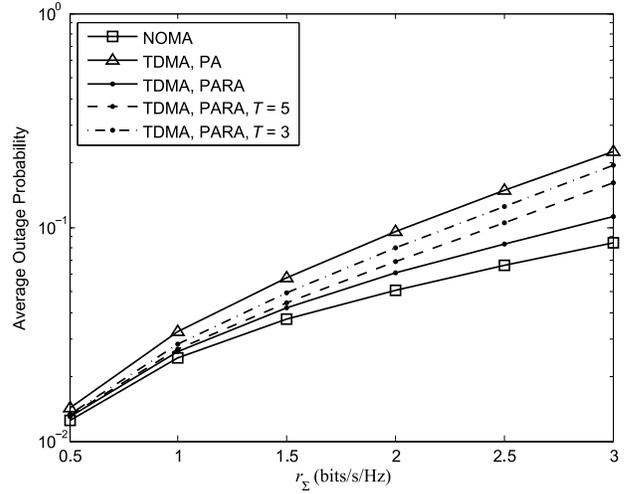}}
\centering\caption{The average outage probability of user $1$ with respect to $r_{\Sigma}$ when $P/N_0=15$ dB and $K=10$.}
\label{fig:w.r.t.rate}
\end{figure}

\begin{figure}[!t]
\centering{\includegraphics[scale=0.65]{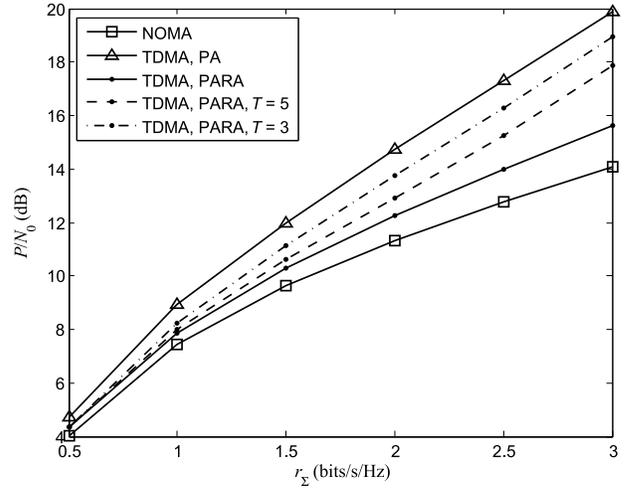}}
\centering\caption{The transmit SNR required by NOMA and TDMA to achieve an average outage probability of $0.1$ for the users when $K=10$.}
\label{fig:w.r.t.rateSNRgain}
\end{figure}

\subsection{User fairness enhancement by NOMA}
\label{sec:SimNoGroup}
To evaluate how does NOMA performs on the average, we simulate the elementary downlink NOMA system for $5000$ times to cover a large number of scenarios; then, for each user, the \emph{average of its outage probability over all simulations} (average outage probability in short) is used as a performance measure. At each time, the simulation parameters $\bm r$, $\bm w$, and $\bm d$ are generated independently. The data rates follow a uniform distribution under the constraint that $\sum_{k=1}^{K}r_k=r_{\Sigma}$, where $r_{\Sigma}$ is the sum of the targeted data rates of the users, which can also be referred to as the targeted system spectral efficiency. The weighting factors follow a uniform distribution over $[0,1]$. \emph{By symmetry of the users}, they will have very close average performance. Hence, we only need to focus on one of the $K$ users in the plots. Here, \emph{we choose user 1}.

\begin{figure*}[!t]
\centering{\includegraphics[scale=0.65]{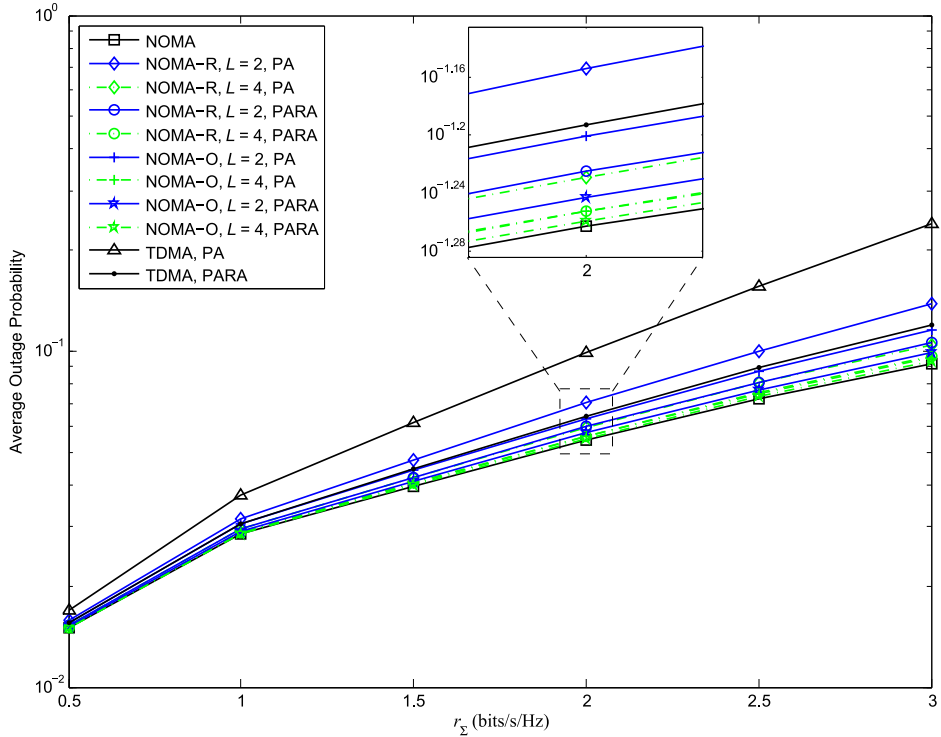}}
\centering\caption{Impact of user grouping algorithm and group size on the performance of NOMA. $P/N_0=15$ dB and $K=8$.}
\label{fig:w.r.t.LevelOpt}
\end{figure*}

Fig. \ref{fig:w.r.t.userNum} shows the average outage probability of user 1 with respect to $K$. It can be seen that the performance gain of NOMA over TDMA increases with $K$ when $K$ is small, and turns to be steady as $K$ becomes large. This can be explained by the fact that SC utilizes the diversity in the channel qualities of the users as a new freedom for potential performance gain \cite{1972BC}. Generally, more users implies more rich channel diversity, while as $K$ becomes large, the increment will be less evident.

Fig. \ref{fig:w.r.t.rate} shows the average outage probability of user 1 at different sum rates. Fig. \ref{fig:w.r.t.rateSNRgain} shows the transmit SNR required by different schemes at different sum rates and at an average outage probability requirement of 0.1 for the users.
It is observed from both the outage performance and power consumption perspective that NOMA always performs better than TDMA, no matter when PA or PARA is adopted by TDMA. Also, we see that the advantage of NOMA becomes increasingly more evident as the data rate increases. This is because NOMA exploits the channel resources more efficiently than OMA, and hence is more beneficial at higher data rate. Note that when the data rate is rather high, the advantage of NOMA will be less evident, because it becomes difficult for all the schemes to support the communication, and both NOMA and OMA will have a bad outage performance. The results are omitted here for save of space.

\subsubsection{Discussion}
From Fig. \ref{fig:w.r.t.rate} and \ref{fig:w.r.t.rateSNRgain}, the performance gain of NOMA over TDMA is limited when resource allocation among the users is available for TDMA. However, the effect of resource allocation in practical systems will be much worse than it theoretically does. Theoretically, the channel resources can be arbitrarily divided among the users, i.e., \emph{continuous resource allocation}. Such an ideal condition is not available in practice. Generally, according to the operating mode of existing communication systems, we can only allocate integer resource blocks to each user, i.e., \emph{discrete resource allocation}. In this sense, resource allocation will result in additional delay, for which the range of resource allocation will be restricted. Even if we divide the original resource block into smaller parts, it is the most possible to divide it into equal parts.

For better understanding, we also show TDMA with discrete time allocation in Fig. \ref{fig:w.r.t.rate} and \ref{fig:w.r.t.rateSNRgain}, where we use $T$ to denote the number of sub-timeslots that each original timeslot can be divided into. Obviously, $T$ reflects the ability in resource allocation of the practical system. As discussed above, discrete resource allocation performs much worse than the theoretically optimal one.

\subsection{Performance and complexity tradeoff}
\begin{figure*}[!t]
\centering{\includegraphics[scale=0.65]{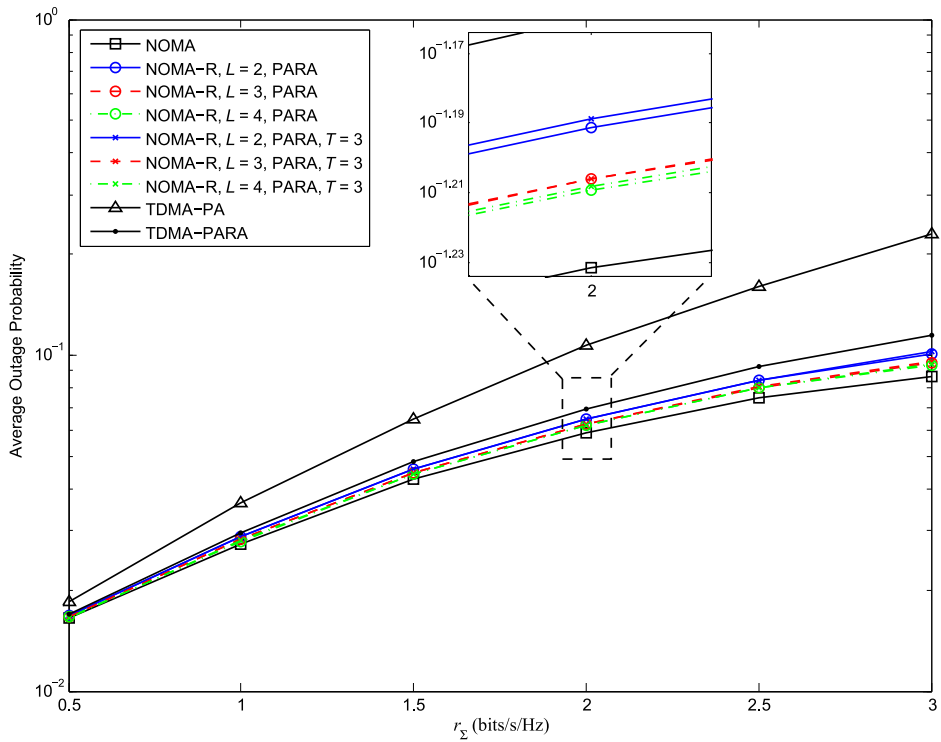}}
\centering\caption{Impact of discrete resource allocation on the performance of NOMA with user grouping. $P/N_0=15$ dB and $K=12$.}
\label{fig:w.r.t.LevelRnd}
\end{figure*}

In this subsection, the case with user grouping is evaluated. The focus is on the complexity and performance tradeoff of NOMA. For notational convenience, we denote by NOMA the case without user grouping, NOMA-O the case with optimal user grouping, and NOMA-R the case with random user grouping.
In addition, we use PA to denote that only inter-group power allocation is adopted by NOMA and use PARA to denote that both inter-group power and resource allocation are adopted by NOMA. The configuration of the parameters $\bm{r}$, $\bm{w}$, and $\bm{d}$ is the same as in the previous subsection. Also, we focus on user $1$ in the plots.

Fig. \ref{fig:w.r.t.LevelOpt} gives the average outage probability of user $1$ with respect to $r_{\Sigma}$ under different user grouping algorithms and group sizes. From Fig. \ref{fig:w.r.t.LevelOpt}, it is observed that both NOMA-R and NOMA-O outperform the corresponding TDMA scheme (either with PA or with PARA), irrespective of the group size $L$. Though NOMA-R performs worse than NOMA-O, the performance gap is small, especially when $L$ is ralatively large and/or resource allocation is adopted. Also, we see that NOMA-R with $L=2$ can achieve a performance very close to that of NOMA when PARA is adopted. Even when resource allocation is unavailable, NOMA-R with $L$ between 2 and 4 can reap a large portion of the performance gain of NOMA (over TDMA with PA).
These obvervations imply that NOMA with rand user grouping and a small group size can achieve a good enough performance. We know that the complexity of inter-group resource allocation in NOMA is the same to that of time allocation in TDMA, hence the additional complexity of user grouping mainly lies in the user grouping algorithm. Since the complexity of random user grouping is negligable compared to that of SIC, it is attractive and also effective to improve the compexity and performance tradeoff of NOMA by user grouping.

In the previous subsection, we have discussed that due to the physical limitations in practical communication systems, resource allocation can only be performed in a discrete manner, which may degrade the effectiveness of resource allocation. Fig. \ref{fig:w.r.t.LevelRnd} illustrates the impact of such restriction on the effectiveness of inter-group resource allocation in a downlink NOMA system. Interestingly, the case with discrete resource allocation and $T=3$ achieves a performance very close to that with the optimal continuous one. This is because in NOMA, the performance of the users in the same group have already been balanced by nonorthogonal transmission with fairness power allocation. As a result, the imbalance in the performance of different user groups becomes less severe, and hence the demand for resource allocation (to enhance user fairness) is much weaker than in a TDMA system.

\section{conclusion}
\label{sec:conclusion}
The outage balancing problem was investigated to achieve the optimal fairness outage performance in a downlink NOMA system when only statistical CSI is available at the transmitter. Both the problems of power allocation and decoding order selection were thoroughly studied and solved analytically. It was proved that the optimal decoding order is determined by the ordering of the weighted average channel gains of the users. The simulation results for both the cases with and without user grouping illustrated that NOMA performs much better than OMA in terms of the fairness outage performance, especially when the practical discrete resource allocation restriction is taken into account. For the case with user grouping, we solved the problems of inter-group power and resource allocation. It was substantiated by simulations that user grouping serves as an effective method to reduce the implementation complexity of NOMA due to SIC. The complexity issue of user grouping  can be well circumvented by using a random user grouping algorithm,  which has negligible complexity when compared with that of SIC. It was shown that NOMA with random user grouping and a small group size (2 to 4 users in each group) can reap most of the performance gain of NOMA.

\appendices
\section{Proof of Theorem \ref{thm:PAF}}
\label{sec:proofofTheorem1}
It has already been proved in Section \ref{sec:PAF} that $\gamma_{\textrm{th}}^{\pi_1}\leq\gamma_{\textrm{th}}^{\pi_2}$, we only need to further prove that $\gamma_{\textrm{th}}^{\pi_m}\leq\gamma_{\textrm{th}}^{\pi_{m+1}}$ for $2\leq m< K$ under the assumption that
\begin{equation}
    \gamma_{\textrm{th}}^{\pi_1}\leq\gamma_{\textrm{th}}^{\pi_2}\leq\cdots \leq\gamma_{\textrm{th}}^{\pi_m}\textrm{.}
    \label{eq:constraint2assump}
\end{equation}
Consider the decoding of any two adjacent signals $x_{\pi_m}$ and $x_{\pi_{m+1}}$, $2\leq m<K$ at an arbitrary user $U_{\pi_n}$, $2\leq n<K$ under the assumption that the constraints in \eqref{eq:constraint1}, \eqref{eq:sumto1}, and \eqref{eq:constraint2assump} are satisfied and the PAFs of all the users other than $U_{\pi_m}$ and $U_{\pi_{m+1}}$ are given. Then, the outage events in decoding $x_{\pi_m}$ and $x_{\pi_{m+1}}$ at $U_{\pi_n}$ can be given by
\begin{IEEEeqnarray}{rCl}
    \mathcal{O}^{\pi_n}_{\pi_m}
    &=&\left\{\bigcup_{j\in\mathcal{K},j\leq m}\gamma_{\pi_j}^{\pi_n}< 2^{r_{\pi_j}}-1 \right\}
    \nonumber\\
    &=&\left\{\bigcup_{j\in\mathcal{K},j\leq m}
    \gamma_{\pi_n}<\gamma_{\textrm{th}}^{\pi_j}\right\}
    \nonumber\\
    &=&\big\{\gamma_{\pi_n}<\gamma_{\textrm{th}}^{\pi_m}\big\}
\label{eq:outevent_pim_at_pim}
\\
    \textrm{and }
    \mathcal{O}^{\pi_{n}}_{\pi_{m+1}}
    &=&\left\{\gamma_{\pi_{n}} < \gamma_{\textrm{th}}^{\pi_{m+1}}\right\}
    \bigcup\mathcal{O}^{\pi_{n}}_{\pi_m}
    \nonumber\\
    &=&\left\{\gamma_{\pi_{n}} < \max\left(\gamma_{\textrm{th}}^{\pi_{m+1}},\gamma_{\textrm{th}}^{\pi_m}\right)\right\},
\label{eq:outevent_pim+1_at_pim+1}
\end{IEEEeqnarray}
respectively. It is worth mentioning that the second and third steps in \eqref{eq:outevent_pim_at_pim} are based on the assumptions stated in \eqref{eq:constraint1} and \eqref{eq:constraint2assump}, respectively.

Based on $\sum_{k\in\mathcal{K}}\alpha_{\pi_k}=1$, $\alpha_I^{\pi_m}=\alpha_{\pi_{m+1}}+\alpha_I^{\pi_{m+1}}$,
and the constraints in \eqref{eq:constraint1} (for $k = m$ and $m+1$), the feasible range of $\alpha_{\pi_{m+1}}$ is obtained as
\begin{equation}
    \alpha_{\pi_{m+1}}\in\left((2^{r_{\pi_{m+1}}}-1)\alpha_I^{\pi_{m+1}}, \frac{1-\sum_{k=1}^{m-1}\alpha_{\pi_k}}{2^{r_{\pi_m}}}-\alpha_I^{\pi_{m+1}}\right).
\end{equation}
Following the same lines as the discussion on the selection of $\alpha_{\pi_2}$ in section \ref{sec:PAF}, it can be prove that for the decoding of $x_{\pi_m}$, $x_{\pi_{m+1}}$, and the following signals (i.e., $x_{\pi_{m+2}}$, $x_{\pi_{m+3}}$, $\cdots$, $x_{\pi_{K}}$), the optimal $\alpha_{\pi_m}$ and $\alpha_{\pi_{m+1}}$ should satisfy $\gamma_{\textrm{th}}^{\pi_m}\leq\gamma_{\textrm{th}}^{\pi_{m+1}}$. Note that the selection of $\alpha_{\pi_m}$ and $\alpha_{\pi_{m+1}}$ does not affect the decoding of the signals prior to $x_{\pi_m}$ (i.e., $x_{\pi_1},x_{\pi_2},\cdots,x_{\pi_{m-1}}$), it can be concluded that the constraint $\gamma_{\textrm{th}}^{\pi_{m+1}}\geq\gamma_{\textrm{th}}^{\pi_m}$ is optimal for the whole system. Since $x_{\pi_m}$ and $x_{\pi_{m+1}}$ are arbitrarily assumed, the optimality of the constraints in \eqref{eq:constraint2} can be proved.

\section{Proof of Theorem \ref{thm:order}}
\label{sec:proofofTheorem2}

Assume optimal power allocation among the users according to Theorem \ref{thm:PAF}. Then from $\gamma_{\textrm{th}}^{\pi_m}\leq\gamma_{\textrm{th}}^{\pi_{m+1}}$ in \eqref{eq:constraint2}, we obtain an upper bound on $\alpha_{\pi_{m+1}}$ as follows
\begin{equation}
    \alpha_{\pi_{m+1}}\leq\frac{2^{r_{\pi_{m+1}}}-1}{2^{r_{\pi_m}}2^{r_{\pi_{m+1}}}-1} \left(1-\bar\alpha\right)\textrm{,}
    \label{eq:original_rephrased}
\end{equation}
where $\bar\alpha=\sum_{k\in\mathcal{K},k\neq m,m+1}\alpha_{\pi_k}$ denotes the sum of the PAFs of all the users except $U_{\pi_m}$ and $U_{\pi_{m+1}}$.
First, we will show that for any feasible values of $\alpha_{\pi_m}$ and $\alpha_{\pi_{m+1}}$, we can find new PAFs $\beta_{\pi_m}$ and $\beta_{\pi_{m+1}}$ for $U_{\pi_m}$ and $U_{\pi_{m+1}}$, respectively, such that the following constraints are satisfied
\begin{equation}
    \beta_{\pi_m}+\beta_{\pi_{m+1}} = 1-\bar\alpha\textrm{,}
\label{eq:conservation}
\end{equation}
\begin{equation}
    \widetilde\gamma_{\textrm{th}}^{\pi_{m+1}}\leq\widetilde{\gamma}_{\textrm{th}}^{\pi_m} \leq\gamma_{\textrm{th}}^{\pi_{m+1}}\textrm{,}
    \label{eq:new}
\end{equation}
where in \eqref{eq:new}, we use the following notations
\begin{align}
    \widetilde{\gamma}_{\textrm{th}}^{\pi_{m+1}}&=
    \frac{2^{r_{\pi_{m+1}}}-1}{\beta_{\pi_{m+1}}-(2^{r_{\pi_{m+1}}}-1)(\alpha_I^{\pi_{m+1}}+\beta_{\pi_m})}\textrm{,}
    \\
    \widetilde{\gamma}_{\textrm{th}}^{\pi_m}&=
    \frac{2^{r_{\pi_m}}-1}{\beta_{\pi_m}-(2^{r_{\pi_m}}-1)\alpha_I^{\pi_{m+1}}}
    \textrm{.}
\end{align}
In fact, the first inequality in \eqref{eq:new} can be satisfied if
\begin{IEEEeqnarray}{rCl}
    \beta_{\pi_{m+1}}&\geq&\left(1-\frac{2^{r_{\pi_m}}-1}{2^{r_{\pi_m}}2^{r_{\pi_{m+1}}}-1}\right) \left(1-\bar\alpha\right),
    \label{eq:new1_rephrased}
\end{IEEEeqnarray}
which is definitely feasible, and the second inequality in \eqref{eq:new} is equivalent to
\begin{equation}
    \beta_{\pi_{m+1}}\leq1-\bar\alpha- \frac{2^{r_{\pi_m}}-1}{2^{r_{\pi_{m+1}}}-1}\alpha_{\pi_{m+1}}\textrm{,}
    \label{eq:new2_rephrased}
\end{equation}
which is feasible without violating the first one considering the precondition in \eqref{eq:original_rephrased}.
The inequalities in \eqref{eq:new} imply that if we use the new PAFs for $x_{\pi_m}$ and $x_{\pi_{m+1}}$, $x_{\pi_m}$ should be decoded after $x_{\pi_{m+1}}$ while the decoding orders of the signals $x_{\pi_{m+2}}$, $x_{\pi_{m+3}}$, $\cdots$, $x_{\pi_{K}}$ will not be affected.
The equality in \eqref{eq:conservation} implies that the PAFs of the other users are not changed. So if we keep the decoding orders of the signals prior to $x_{\pi_m}$ (i.e., $x_{\pi_{m+2}}$, $x_{\pi_{m+3}}$, $\cdots$, $x_{\pi_{K}}$) unchanged, the outage probabilities of all the users other than $U_{\pi_m}$ and $U_{\pi_{m+1}}$ will not be affected by the selection of the new PAFs.

Now suppose that we have exchanged the decoding orders of $x_{\pi_m}$ and $x_{\pi_{m+1}}$ by using the new PAFs $\beta_{\pi_m}$ and $\beta_{\pi_{m+1}}$ that satisfy \eqref{eq:new1_rephrased} and \eqref{eq:new2_rephrased}, and the decoding orders of the other user signals are unchanged. Use $\mathcal{\widetilde{O}}^{\pi_m}_{\pi_m}$ and $\mathcal{\widetilde{O}}^{\pi_{m+1}}_{\pi_{m+1}}$ to denote the outage events of user $U_{\pi_m}$ and $U_{\pi_{m+1}}$, respectively, under the new decoding order. Then, we have
\begin{IEEEeqnarray}{rCl}
    \mathcal{\widetilde{O}}^{\pi_{m+1}}_{\pi_{m+1}}
    &=&\left\{\gamma_{\pi_{m+1}}< \widetilde{\gamma}_{\textrm{th}}^{\pi_{m+1}}
    \right\}\bigcup\mathcal{O}^{\pi_{m+1}}_{\pi_{m-1}}
    \nonumber\\
    &=&\left\{\gamma_{\pi_{m+1}}< \max\left(\widetilde{\gamma}_{\textrm{th}}^{\pi_{m+1}},
    \gamma_{\textrm{th}}^{\pi_{m-1}}\right)\right\}\textrm{,}
    \label{eq:outevent_pim+1_at_pim+1_tilde}
\\
    \mathcal{\widetilde{O}}^{\pi_m}_{\pi_m}
    &=&\left\{\gamma_{\pi_m}< \widetilde{\gamma}_{\textrm{th}}^{\pi_m}\right\} \bigcup\widetilde{\mathcal{O}}^{\pi_m}_{\pi_{m+1}}
    \nonumber\\
    &=&\left\{\gamma_{\pi_m}< \widetilde{\gamma}_{\textrm{th}}^{\pi_m}\right\} \bigcup
    \left\{\gamma_{\pi_{m}}< \max\left(\widetilde{\gamma}_{\textrm{th}}^{\pi_{m+1}},
    \gamma_{\textrm{th}}^{\pi_{m-1}}\right)\right\}
    \nonumber\\
    &=&\left\{\gamma_{\pi_m}< \max\left(\widetilde{\gamma}_{\textrm{th}}^{\pi_m},
    \gamma_{\textrm{th}}^{\pi_{m-1}}\right)\right\}
    \textrm{.}
    \label{eq:outevent_pim_at_pim_tilde}
\end{IEEEeqnarray}
Note that in the second step of \eqref{eq:outevent_pim_at_pim_tilde}, we have expanded  $\widetilde{\mathcal{O}}^{\pi_m}_{\pi_{m+1}}$ which can be similarly treated as $\mathcal{\widetilde{O}}^{\pi_{m+1}}_{\pi_{m+1}}$ in \eqref{eq:outevent_pim+1_at_pim+1_tilde}, and in the third step, we have used the inequalities in \eqref{eq:new}. From \eqref{eq:outevent_pim+1_at_pim+1_tilde} and \eqref{eq:outevent_pim_at_pim_tilde}, we obtain the weighted success probabilities of $U_{\pi_{m}}$ and $U_{\pi_{m+1}}$ under the new decoding order as follows
\begin{align}
    \left(1-\textrm{Pr}\left\{\mathcal{\widetilde{O}}^{\pi_m}_{\pi_m}\right\}\right)^{w_{\pi_m}}
    &=\exp\left(-\frac{\max\left(\widetilde{\gamma}_{\textrm{th}}^{\pi_m},\gamma_{\textrm{th}}^{\pi_{m-1}}\right)} {\Gamma_{\pi_m}/w_{\pi_m}}\right)
    \textrm{,}
    \label{eq:outprob_pim_exchanged}
    \\
    \left(1-\textrm{Pr}\left\{\mathcal{\widetilde{O}}^{\pi_{m+1}}_{\pi_{m+1}}\right\}\right)^{w_{\pi_{m+1}}}
    &=\exp\left(-\frac{\max\left(\widetilde{\gamma}_{\textrm{th}}^{\pi_{m+1}},\gamma_{\textrm{th}}^{\pi_{m-1}}\right)} {\Gamma_{\pi_{m+1}}/w_{\pi_{m+1}}}\right)
    \textrm{.}
\end{align}
Under the original decoding order, the weighted success probabilities of $U_{\pi_{m}}$ and $U_{\pi_{m+1}}$ are given by
\begin{align}
    \left(1-\textrm{Pr}\left\{\mathcal{O}^{\pi_m}_{\pi_m}\right\}\right)^{w_{\pi_m}}
    &=\exp\left(-\frac{\gamma_{\textrm{th}}^{\pi_m}} {\Gamma_{\pi_m}/w_{\pi_m}}\right)\textrm{,}
    \\
    \left(1-\textrm{Pr}\left\{\mathcal{O}^{\pi_{m+1}}_{\pi_{m+1}}\right\}\right)^{w_{\pi_{m+1}}}
    &=\exp\left(-\frac{\gamma_{\textrm{th}}^{\pi_{m+1}}}{\Gamma_{\pi_{m+1}}/w_{\pi_{m+1}}}\right)
    \textrm{.}
    \label{eq:outprob_pim+1}
\end{align}
With the facts that  $\widetilde{\gamma}_{\textrm{th}}^{\pi_{m+1}}\leq\widetilde{\gamma}_{\textrm{th}}^{\pi_m} \leq\gamma_{\textrm{th}}^{\pi_{m+1}}$
and $\gamma_{\textrm{th}}^{\pi_{m-1}}\leq\gamma_{\textrm{th}}^{\pi_m}\leq\gamma_{\textrm{th}}^{\pi_{m+1}}$, it can be proved that when $\Gamma_{\pi_m}/w_{\pi_m}>\Gamma_{\pi_{m+1}}/w_{\pi_{m+1}}$,
\begin{multline}
    \min\left(\left(1-\textrm{Pr}\left\{\mathcal{\widetilde{O}}^{\pi_m}_{\pi_m}\right\}\right)^{w_{\pi_m}},
    \left(1-\textrm{Pr}\left\{\mathcal{\widetilde{O}}^{\pi_{m+1}}_{\pi_{m+1}}\right\}\right)^{w_{\pi_{m+1}}} \right)
    \\
    \geq\min\left(\left(1-\textrm{Pr}\left\{\mathcal{O}^{\pi_m}_{\pi_m}\right\}\right)^{w_{\pi_m}}, \left(1-\textrm{Pr}\left\{\mathcal{O}^{\pi_{m+1}}_{\pi_{m+1}}\right\}\right)^{w_{\pi_{m+1}}} \right).
\end{multline}
In other words, if $\Gamma_{\pi_m}/w_{\pi_m}>\Gamma_{\pi_{m+1}}/w_{\pi_{m+1}}$, the minimum of the weighted success probabilities of $U_{\pi_m}$ and $U_{\pi_{m+1}}$ after exchanging the decoding orders of their signals is greater than or equal to that under the original decoding order.

\section{Proof of Proposition \ref{pro:equalWSP}}
\label{sec:proofofProposition1}

We will prove this by controdiction. Suppose that not all $\gamma_{\textrm{th}}^{\pi_k^o}/\Gamma_{\pi_k^o}\cdot w_{\pi_k^o}$ are equal at the optimal solution and denote by $\mathcal{K}'$ the set of the indices such that $\gamma_{\textrm{th}}^{\pi_{k'}^o}/\Gamma_{\pi_{k'}^o}\cdot w_{\pi_{k'}^o}=\max_{k\in\mathcal{K}}\gamma_{\textrm{th}}^{\pi_k^o}/\Gamma_{\pi_k^o}\cdot w_{\pi_k^o}$ for any $k'\in\mathcal{K}'$. Now, if we scale up all $\alpha_{\pi_{k'}^o}$, $k'\in\mathcal{K}'$ by $\delta'>1$ and scale down all $\alpha_{\pi_k^o}$, $k\in\mathcal{K}\setminus\mathcal{K}'$ by $\delta=\frac{\sum_{k\in\mathcal{K}\setminus\mathcal{K}'}\alpha_{\pi_k^o} - \sum_{k'\in\mathcal{K}'}\alpha_{\pi_{k'}^o}(1-\delta')}{\sum_{k\in\mathcal{K}-\mathcal{K}'}\alpha_{\pi_k^o}}<1$, $\gamma_{\textrm{th}}^{\pi_{k'}^o}/\Gamma_{\pi_{k'}^o}\cdot w_{\pi_{k'}^o}$ for all $k'\in\mathcal{K}'$ can be decreased while $\gamma_{\textrm{th}}^{{\pi_k^o}}/\Gamma_{{\pi_k^o}}\cdot w_{{\pi_k^o}}$, $k\in\mathcal{K}\setminus\mathcal{K}'$ will increase, which is straightforward from the definition of $\gamma_{\textrm{th}}$ (see \eqref{eq:outthresh}). The value of $\delta$ ensures that the constraint in (\ref{eq:min-maxop-reform2}b) is satisfied. In addition, it can proved that if $\delta'$ is small enough, the constraints in (\ref{eq:min-maxop-reform2}c) will not be violated, and the maximum of $\gamma_{\textrm{th}}^{{\pi_k^o}}/\Gamma_{{\pi_k^o}}\cdot w_{{\pi_k^o}}$, $k\in\mathcal{K}\setminus\mathcal{K}'$ can be smaller than the minimum of $\gamma_{\textrm{th}}^{\pi_{k'}^o}/\Gamma_{\pi_{k'}^o}\cdot w_{\pi_{k'}^o}$, $k'\in\mathcal{K}'$.
The proof of the second part is stratightforward, hence we focus on the proof of the first part in the following.

\begin{figure*}[!t]
\normalsize
\setcounter{MYtempeqncnt}{\value{equation}}
\setcounter{equation}{52}
\begin{IEEEeqnarray}{rlCl}\label{eq:tildeoutthresh}
  &\frac{2^{r_{\pi_k^o}}-1}{\delta\alpha_{\pi_k^o}-\left(2^{r_{\pi_k^o}}-1\right)
   \left(\sum_{j>k,j\in\mathcal{K}\setminus\mathcal{K'}}\delta\alpha_{\pi_j^o}
   +\sum_{j>k,j\in\mathcal{K'}}\delta'\alpha_{\pi_j^o}\right)}
   \quad&\textrm{, for}\:\:&k\in\mathcal{K},
     \IEEEyesnumber\IEEEyessubnumber
     \label{eq:tildeoutthreshK}
  \\*[-0.1\normalbaselineskip]
  \smash{\tilde{\gamma}_{\textrm{th}}^{\pi_k^o}=\left\{
            \IEEEstrut[11\jot]
         \right.}\nonumber
  \\*[-0.2\normalbaselineskip]
  &\frac{2^{r_{\pi_k^o}}-1}{\delta'\alpha_{\pi_k^o}-\left(2^{r_{\pi_k^o}}-1\right)
   \left(\sum_{j>k,j\in\mathcal{K}\setminus\mathcal{K'}}\delta\alpha_{\pi_j^o}
   +\sum_{j>k,j\in\mathcal{K'}}\delta'\alpha_{\pi_j^o}\right)}
         &\textrm{, for}\:\:&k\in\mathcal{K'}.
     \IEEEyessubnumber
     \label{eq:tildeoutthreshK'}
\end{IEEEeqnarray}
\setcounter{equation}{\value{MYtempeqncnt}}
\hrulefill
\vspace*{0pt}
\end{figure*}
\addtocounter{equation}{1}

Denote by $\tilde{\gamma}_{\textrm{th}}^{\pi_k^o}$ the value of $\gamma_{\textrm{th}}^{\pi_k^o}$ after scaling up or down the PAFs as stated in the above. Then, according to \eqref{eq:outthresh}, the expression of $\tilde{\gamma}_{\textrm{th}}^{\pi_k^o}$ is given in \eqref{eq:tildeoutthresh}; on top of the next page. First, we prove that there exists a $\delta'>1$ satisfying $\tilde{\gamma}_{\textrm{th}}^{\pi_1^o}>0$.
With the fact that $\gamma_{\textrm{th}}^{\pi_1^o}>0$, it is directly from \eqref{eq:outthresh} and \eqref{eq:tildeoutthreshK'} that $\tilde{\gamma}_{\textrm{th}}^{\pi_1^o}>0$ is always satisfied if $1\in\mathcal{K'}$. While if $1\in\mathcal{K}\setminus\mathcal{K'}$, $\tilde{\gamma}_{\textrm{th}}^{\pi_1^o}>0$ can be satisfied only if we choose a small enough $\delta'$ ($>1$), since $\gamma_{\textrm{th}}^{\pi_1^o}>0$. Use $\delta'_1$ ($>1$) to denote the upper bound on the value of $\delta'$ that satisfies $\tilde{\gamma}_{\textrm{th}}^{\pi_1^o}>0$.
Then, we prove that there exists a $\delta'\in\left(1,\delta'_1\right]$ satisfying $\tilde{\gamma}_{\textrm{th}}^{\pi_k^o}\leq\tilde{\gamma}_{\textrm{th}}^{\pi_{k+1}^o}$ for the following four cases,
\begin{itemize}
  \item Case 1: $k\in\mathcal{K}\setminus\mathcal{K'}$, $k+1\in\mathcal{K}\setminus\mathcal{K'}$, and $k<K$;
  \item Case 2: $k\in\mathcal{K'}$, $k+1\in\mathcal{K'}$, and $k<K$;
  \item Case 3: $k\in\mathcal{K'}$, $k+1\in\mathcal{K}\setminus\mathcal{K'}$, and $k<K$;
  \item Case 4: $k\in\mathcal{K}\setminus\mathcal{K'}$, $k+1\in\mathcal{K'}$, and $k<K$.
\end{itemize}
It can be proved that $\tilde{\gamma}_{\textrm{th}}^{\pi_k^o}\leq\tilde{\gamma}_{\textrm{th}}^{\pi_{k+1}^o}$ is always satisfied for the first three cases by simply comparing the expressions of $\tilde{\gamma}_{\textrm{th}}^{\pi_k^o}$ and $\tilde{\gamma}_{\textrm{th}}^{\pi_{k+1}^o}$ with the fact that $\gamma_{\textrm{th}}^{\pi_k^o}\leq\gamma_{\textrm{th}}^{\pi_{k+1}^o}$. For Case 4, we have the following relation of equivalence
\begin{align}
    \tilde{\gamma}_{\textrm{th}}^{\pi_k^o}\leq\tilde{\gamma}_{\textrm{th}}^{\pi_{k+1}^o}
    \Longleftrightarrow
    \frac{\alpha_{\pi_k^o}}{2^{r_{\pi_k^o}}-1}-\frac{\delta'}{\delta}\alpha_{\pi_{k+1}^o}
    \geq
    \frac{\delta'}{\delta}\frac{\alpha_{\pi_{k+1}^o}}{2^{r_{\pi_{k+1}^o}}-1},
    \label{eq:outthresh<tilde}
\end{align}
which can be easily obtained from \eqref{eq:tildeoutthresh}. From $\gamma_{\textrm{th}}^{\pi_k^o}\leq\gamma_{\textrm{th}}^{\pi_{k+1}^o}$, we know that
\begin{equation}
    \frac{\alpha_{\pi_k^o}}{2^{r_{\pi_k^o}}-1}-\alpha_{\pi_{k+1}^o}
    \geq
    \frac{\alpha_{\pi_{k+1}^o}}{2^{r_{\pi_{k+1}^o}}-1}.
    \label{eq:outthresh<}
\end{equation}
Obviously, if the condition in \eqref{eq:outthresh<} is satisfied with inequaltiy, we can find a $\delta'_2>1$ such that \eqref{eq:outthresh<tilde} is satisfied if $\delta'\leq\delta'_2$. It can be seen that the condition in \eqref{eq:outthresh<} can only be satisfied with inequaltiy in Case 4. Since if not so, $\gamma_{\textrm{th}}^{\pi_k^o}$ and $\gamma_{\textrm{th}}^{\pi_{k+1}^o}$ will be equal; then from the fact that $\Gamma_{\pi_k^o}/w_{\pi_k^o}\leq\Gamma_{\pi_{k+1}^o}/w_{\pi_{k+1}^o}$,
$\gamma_{\textrm{th}}^{\pi_k^o}/\Gamma_{\pi_k^o}\cdot w_{\pi_k^o}$ will be greater than or equal to $\gamma_{\textrm{th}}^{\pi_{k+1}^o}/\Gamma_{\pi_{k+1}^o}\cdot w_{\pi_{k+1}^o}$, which violates the fact that $k\in\mathcal{K}\setminus\mathcal{K'}$ and $k+1\in\mathcal{K'}$ in Case 4. From the above discussions, if we choose $\delta'$ in the range of $\left(1,\min\left(\delta'_1,\delta'_2\right)\right]$, the constraint in (\ref{eq:min-maxop-reform2}c) will be satisfied.

It follows that if we choose a suitable $\delta'$, $\max_{k\in\mathcal{K}} \gamma_{\textrm{th}}^{\pi_k^o}/\Gamma_{\pi_k^o}\cdot w_{\pi_k^o}$ can be further decreased, which contradicts the optimality assumption.

\section{Proof of Proposition \ref{pro:convexityfg}}
\label{sec:proofofconvexity}

We first consider the first derivation of $f_g(t_g)$, which is given by
\begin{align}
   f_g'(t_g)=&\frac{2^{r_{g_1}/t_{g}}-1-2^{r_{g_1}/t_{g}}\ln2r_{g_1}/t_{g}}{\Gamma_{g_1}/w_{g_1}}
    \nonumber\\
    &+\:\frac{2^{r_{g_2}/t_{g}}-1-2^{r_{g_2}/t_{g}}\ln2r_{g_2}/t_{g}}{\Gamma_{g_2}/w_{g_2}}2^{r_{g_1}/t_{g}}
    \nonumber\\
    &-\:\frac{2^{r_{g_2}/t_{g}}-1}{\Gamma_{g_2}/w_{g_2}}2^{r_{g_1}/t_{g}}\ln2r_{g_1}/t_{g}
    \nonumber\\
    &\cdots
    \nonumber\\
    &+\:\frac{2^{r_{g_l}/t_{g}}-1-2^{r_{g_l}/t_{g}}\ln2r_{g_l}/t_{g}}{\Gamma_{g_l}/w_{g_l}}2^{\sum_{j=1}^{l-1}r_{g_j}/t_{g}}
    \nonumber\\
    &-\:\frac{2^{r_{g_l}/t_{g}}-1}{\Gamma_{g_l}/w_{g_l}}2^{\sum_{j=1}^{l-1}r_{g_j}/t_{g}}\ln2\sum_{j=1}^{l-1}r_{g_j}/t_{g}
    \nonumber\\
    &\cdots
    \nonumber\\
    &+\:\frac{2^{r_{g_L}/t_{g}}-1-2^{r_{g_L}/t_{g}}\ln2r_{g_L}/t_{g}}{\Gamma_{g_L}/w_{g_l}}2^{\sum_{j=1}^{L-1}r_{g_j}/t_{g}}
    \nonumber\\
    &-\:\frac{2^{r_{g_L}/t_{g}}-1}{\Gamma_{g_L}/w_{g_L}}2^{\sum_{j=1}^{L-1}r_{g_j}/t_{g}}\ln2\sum_{j=1}^{L-1}r_{g_j}/t_{g}.
    \label{eq:derivationfgnoncompact}
\end{align}
After some algebraic manipulations, the expression in \eqref{eq:derivationfgnoncompact} can be rephrased as in \eqref{eq:fgderivate}.

Since $x\ln x - x$ strictly increases with $x$ when $x\geq1$,  and for any $l \in\mathcal{L}= \{1,2,\cdots,L\}$, $x_l>1$ and $x_l$ strictly decreases with $t_g$ when $t_g>0$. It is then clear that each term $x_g^l\ln x_g^l - x_g^l$, $l\in\mathcal{L}$ is a strictly monotone decreasing function of $t_g$ when $t_g>0$. In addition, it is always true that $w_{g_l}/\Gamma_{g_l}>0$ for $l\in\mathcal{L}$ and $w_{g_1}/\Gamma_{g_1}\geq w_{g_2}/\Gamma_{g_2}\geq\cdots\geq w_{g_L}/\Gamma_{g_L}$. Hence, $f_g'(t_g)$ is a strictly monotone increasing function of $t_g$ when $t_g>0$.

With the knowledge that $f_g'(t_g)$ strictly increases with $t_g$ when $t_g>0$, it is intuitive that the second derivate of $f_g(t_g)$ will be positive when $t_g>0$. However, for completeness, we still give the second derivate of $f_g(t_g)$ as follows
\begin{align}
    f_g''(t_g)=
    &\sum_{l=1}^{L-1}\left(\frac{w_{g_l}}{\Gamma_{g_l}}-\frac{w_{g_{l+1}}}{\Gamma_{g_{l+1}}}\right)
    \left(x_g^l(\ln x_g^l)^2/t_g\right)
    \nonumber\\
    &+\frac{w_{g_L}}{\Gamma_{g_L}}\left(x_g^L (\ln x_g^L)^2 /t_g\right), g \in\mathcal{G}
\end{align}
It is obvious that $f_g''(t_g)>0$ when $t_g>0$, which proves the strict convexity of $f_g(t_g)$ with respect to $t_g$ when $t_g>0$.

%\begin{thebibliography}{1}
%\end{thebibliography}
\bibliographystyle{IEEEtran} \bibliography{IEEEabrv,RBC}

\end{document}